\documentclass[11pt]{article}
\usepackage{amsmath}
\usepackage{graphicx}

\numberwithin{equation}{section}

\begin{document}
\baselineskip=15pt
\begin{titlepage}
\begin{flushright}
{\small KYUSHU-HET-89}\\[-1mm] hep-ph/0601152
\end{flushright}
\begin{center}
\vspace*{11mm}

{\Large\bf%
Minimal archi-texture for neutrino mass matrices
}\vspace*{8mm}

Atsushi Watanabe and Koichi Yoshioka
\vspace*{2mm}

{\it Department of Physics, Kyushu University, Fukuoka 812-8581, Japan}

\vspace*{5mm}

{\small (January, 2006)}
\end{center}
\vspace*{5mm}

\begin{abstract}\noindent%
The origin of the observed masses and mixing angles of quarks and
leptons is one of imperative subjects in and beyond the standard
model. Toward a deeper understanding of flavor structure, we
investigate in this paper the minimality of fermion mass (Yukawa)
matrices in unified theory. That is, the simplest matrix form is
explored in light of the current experimental data for quarks and
leptons, including the recent measurements of quark CP violation and
neutrino oscillations. Two types of neutrino mass schemes are
particularly analyzed; (i) Majorana masses of left-handed neutrinos
with unspecified mechanism and (ii) Dirac and Majorana masses
introducing three right-handed neutrinos. As a result, new classes of
neutrino mass matrices are found to be consistent to the low-energy
experimental data and high-energy unification hypothesis. For
distinctive phenomenological implications of the minimal fermion mass
textures, we discuss flavor-violating decay of charged leptons, the
baryon asymmetry of the universe via thermal leptogenesis,
neutrino-less double beta decay, and low-energy leptonic CP violation. 
\end{abstract}
\end{titlepage}

\section{Introduction}
While the standard model successfully describes the gauge interactions
of quarks and leptons, the origin of fermion masses and mixing angles
is not yet understood. The standard model treats these quantities as
free parameters, namely the coupling constants of Yukawa
interactions. In the three-generation framework, Yukawa couplings are
given by complex $3\times3$ matrices in the generation space. A
natural way to remedy the problem is to assume that some physics at
high-energy scale dynamically controls Yukawa couplings and brings us
the observed patterns of masses and mixing angles. 

Along this line, various forms of Yukawa matrices have been discussed
in the literature both for quarks and leptons. For the quark sector,
possible small numbers of non-vanishing matrix elements have been
examined~\cite{quarkM} whether they are consistent with the
experimental results, and the obtained forms of matrices have provided 
us clues to find symmetry or dynamics behind the Yukawa
interactions. There have also been many studies about possible
non-vanishing elements of lepton Yukawa matrices. In particular,
neutrino mass matrices are classified from various
standpoints~\cite{leptonM} and that gives new insights about neutrino
models to produce the observed data. Simultaneous treatments of quark
and lepton Yukawa matrices lead to the investigations in unified
theory based on the gauge groups such 
as $SU(4)$~\cite{PS}, $SU(5)$~\cite{SU5}, $SO(10)$~\cite{SO10}, 
and $E_6$~\cite{E6}. An interesting point of unified theory is that 
the unification of standard model gauge groups implies non-trivial
relations among various coupling constants including Yukawa
couplings. The cerebrated bottom/tau unification~\cite{btau} and other
theoretical predictions for masses and mixing
angles~\cite{GUTrela} suggest that the unification hypothesis
might provide us a deeper understanding of Yukawa matrices.

The recent experimental data on fermion masses and mixing angles,
however, gives us some questions about the unification hypothesis. In
particular, the neutrino oscillation experiments have been bringing
out the structure of leptonic generation mixing which indicates
some kind of quark-lepton asymmetry, that is, large lepton mixing
against small quark mixing. In the light of quark-lepton unification,
that seems somewhat unnatural since the generation structure of quarks
is often correlated to that of leptons in unified theory. A possible
solution to avoid such an intuitive result is to assume fermion mass
matrices which are non-symmetric in the generation
space~\cite{lopsided}. This idea may have dynamical verification,
e.g.\ in $SU(5)$ unification where right-handed down-type quarks are
combined with left-handed charged leptons. Consequently, since their
mass matrices are generally left-right asymmetric, large leptonic
mixing is derived without disturbing quark mixing angles, and can be
consistent with unified gauge symmetry.\footnote{Leptonic mixing
angles might be enhanced by integrating out heavy fields such as
right-handed neutrinos while keeping correlative quark and lepton
Yukawa couplings~\cite{enhance}.} Given this fact, it is meaningful to
explore general, not necessarily symmetric, forms of 
mass (Yukawa) matrices for quarks and leptons including the neutrino
sector. With a throughout analysis in the light of the recent
experimental data of quarks as well as leptons, new insights into the
origin of generation structure are expected to be obtained.

In a previous work~\cite{asym}, we have analyzed in detail the quark
sector where the down-quark mass matrices are of asymmetrical
forms. Based on the minimality principle, namely with the smallest
number of non-vanishing matrix elements, we have found 6 patterns of
up and down-quark matrices which are entirely accommodated to the
observed values of quark mass and mixing parameters. In addition, they
are capable of leading to large leptonic mixing 
via $SU(5)$ relation. In this paper, we complete the analysis by
exploring lepton (neutrino) mass matrices with the minimal number of
independent mass matrix elements. We consider two types of schemes for
neutrino masses. The first is to analyze light Majorana masses of the
three-generation left-handed neutrinos, that is, $3\times 3$ Majorana
mass matrices. The second scheme introduces three right-handed
neutrinos which have large Majorana masses and also Dirac masses with
the left-handed ones. In this scheme, we are lead to simultaneously
analyze the minimality of Dirac and Majorana mass matrices. The
purposes of this paper are to present the simplest forms of lepton
mass matrices with which the 6 quark mass textures previously found
are compatible and also to study some phenomenology related to the
obtained forms of lepton mass matrices.

The paper is organized as follows. In Section 2, we describe the
standard form of Yukawa interactions of quarks and leptons, and
introduce several parameters and their experimental values needed in
later analysis. In Section 3, we revisit the quark mass textures found
in the previous work to examine the discrepancy between down-type
quarks and charged leptons, especially for the first and second
generations. Based on these results, the minimality is explored for
two types of neutrino mass models; Majorana mass matrices of
left-handed neutrinos (in Section 4) and Dirac and Majorana masses
introducing right-handed neutrinos (in Section 5). Section 6 is
devoted to analyzing some phenomenological issues of the lepton mass
textures found in the previous sections. Section 7 summarizes our
results.

\section{Formulation and input parameters}
\label{sec:exp}
We first present the standard-model formulation of Yukawa interactions
for quarks and leptons and the observed values of mass and mixing
parameters, which are used as input parameters. The Yukawa terms
invariant under the standard-model gauge transformations are given by
\begin{equation}
  -{\cal L} \,=\, \overline Q_i (Y_u)_{ij} u_R{}_j H^* 
  +\overline Q_i (Y_d)_{ij} d_R{}_j H 
  +\overline{e_R}{}_i(Y_e)_{ij} L_j H^* \,+\text{h.c.},
\end{equation}
where $Q$ and $L$ denote the electroweak doublets of left-handed
quarks and 
leptons, $Q=(u_L,d_L)^{\rm T}$ and $L=(\nu_L,e_L)^{\rm T}$, 
and $u_R$, $d_R$, $e_R$ are the right-handed up-type quarks, 
down-type quarks and charged leptons, respectively. The Yukawa
coupling constants $Y_u$, $Y_d$ and $Y_e$ are 
complex-valued $3\times3$ matrices ($i$ and $j$ are generation indices
running over 1 to 3), and $H$ is the Higgs doublet. There is no
renormalizable Yukawa term for neutrinos in the standard model. After
the electroweak symmetry breaking, these Yukawa interactions lead to
the following quark and lepton mass terms:
\begin{gather}
  -{\cal L} \,=\, \overline{u_L}{}_i (M_u)_{ij} u_R{}_j 
  +\overline{d_L}{}_i (M_d)_{ij} d_R{}_j 
  +\overline{e_R}{}_i (M_e)_{ij} e_L{}_j \,+\text{h.c.}, \\[2mm]
  \quad (M_u)_{ij}=(Y_u)_{ij}v, \qquad (M_d)_{ij}=(Y_d)_{ij}v,
  \qquad (M_e)_{ij}=(Y_e)_{ij} v, \nonumber
\end{gather}
where $v$ is the vacuum expectation value of the neutral component of
Higgs scalar $H$. In addition to the above mass terms for charged
fermions, one may consider Majorana mass term for left-handed
neutrinos, taking into account the electroweak symmetry breaking. That
is parametrized as
\begin{equation}
  -{\cal L} \,=\, 
  \frac{1}{2}\nu_L{}_i^{\rm T}(M_L)_{ij}C \nu_L{}_j+\textrm{h.c.},
  \label{ML}
\end{equation}
where $C$ is the charge conjugation matrix. The Majorana mass matrix
is generally complex-valued and has a symmetry 
property ${M_L}^{\!\rm T}=M_L$. Throughout this paper, we assume that
light neutrinos are Majorana particles.

The generation mixings are described by the Cabibbo-Kobayashi-Maskawa
(CKM) matrix~\cite{CKM} for quarks and the Maki-Nakagawa-Sakata (MNS)
matrix~\cite{MNS} for leptons, each of which consists of two unitary
matrices
\begin{equation}
  V_{\rm CKM} \,=\, V_{uL}^\dagger V_{dL}^{}, \qquad
  V_{\rm MNS} \,=\, V_{eL}^\dagger V_\nu^{}.
\end{equation}
These unitary matrices diagonalize the mass matrices of quarks and
leptons as
\begin{alignat}{4}
  M_u &\,=\, V_{uL}^{} M_u^D V_{uR}^\dagger, & \qquad
  M_d &\,=\, V_{dL}^{} M_d^D V_{dR}^\dagger, \\
  M_e &\,=\, V_{eR}^{} M_e^D V_{eL}^\dagger, & \qquad
  M_L &\,=\, V_\nu^* M_L^D V_\nu^\dagger,
\end{alignat}
where $V_{uR}$, $V_{dR}$ and $V_{eR}$ are unitary matrices which can
be removed by unitary rotations of right-handed fermions. The diagonal
elements of $M_u^D$, $M_d^D$, and $M_e^D$ correspond to the
experimentally observed mass eigenvalues; 
$M_u^D={\rm diag}\,(m_u,m_c,m_t)$, $M_d^D={\rm diag}\,(m_d,m_s,m_b)$, 
and $M_e^D={\rm diag}\,(m_e,m_\mu,m_\tau)$, respectively, where all
the eigenvalues are made real and positive. As for the neutrino
masses, $M_L^D={\rm diag}\,(m_1,m_2,m_3)$ and the squared mass
differences are defined 
as $\Delta m^2_{ij}\equiv m^2_i-m^2_j$ ($i,j=1,2,3$), which are the
observables in neutrino oscillation experiments.

The current masses for quarks and charged leptons at the $Z$-boson
mass scale are evaluated including various effects such as QCD
coupling effects~\cite{PDG,FK}
\begin{equation}
\begin{array}{lll}
  m_u \,=\, 0.000975-0.00260, & m_d \,=\, 0.00260-0.00520, & 
  m_e \,=\, 0.00048685, \\
  m_c \,=\, 0.598-0.702, & m_s \,=\, 0.0520-0.0845, & 
  m_\mu \,=\, 0.10275, \\
  m_t \,=\, 170-180, & m_b \,=\, 2.83-3.04, & m_\tau \,=\, 1.7467,
\end{array}
\label{massexp}
\end{equation}
in GeV unit. For the charged lepton masses, we have neglected errors
and will refer to the above fixed values because the errors of charged
lepton masses are very small and do not affect the analysis of mass
matrix forms.

The physical mixing parameters in the quark sector are 3 mixing angles
and 1 complex phase in the CKM matrix. The measured magnitudes of the
CKM matrix elements are~\cite{PDG}
\begin{equation}
  |V_{\rm CKM}|=\left|\begin{pmatrix}
    V_{ud} & V_{us} & V_{ub} \\
    V_{cd} & V_{cs} & V_{cb} \\
    V_{td} & V_{ts} & V_{tb}
  \end{pmatrix}\right|\,= \begin{pmatrix}
    0.9739-0.9751 & 0.221-0.227 & 0.0029-0.0045 \\
    0.221-0.227 & 0.9730-0.9744 & 0.039-0.044 \\
    0.0048-0.014 & 0.037-0.043 & 0.9990-0.9992
  \end{pmatrix}.
  \label{CKMexp}
\end{equation}
From various observations of CP-violating processes in the quark
sector, the rephasing-invariant measure of CP violation~\cite{Jcp} is
given by
\begin{equation}
  J_{\rm CP} \,=\, (2.88 \pm 0.33) \times 10^{-5}.
\end{equation}
In addition, the experimental study of the $B$-meson decay to
charmoniums indicates~\cite{CP-B}
\begin{equation}
  \sin 2\phi_1/\beta \,=\, 0.726\pm 0.037,
  \label{sin2b}
\end{equation}
where $\phi_1=\beta$ is one of the angles of the unitary triangle for
the $B$-meson system which is defined 
as $\phi_1=\beta\equiv\arg\,(V_{cd}^*V_{cb}^{}V_{td}^{}V_{tb}^*)$. 

The physical mixing parameters in the lepton sector are 3 mixing
angles and 3 complex phases in the MNS matrix for Majorana neutrinos.
We use the following parametrization for the mixing matrix:
\begin{equation} 
  V_{\rm MNS} \,=\,\begin{pmatrix}
    c_{12}c_{13} & s_{12}c_{13} & s_{13}e^{-i\delta} \\
    -s_{12}c_{23}-c_{12}s_{23}s_{13}e^{i\delta}
    & c_{12}c_{23}-s_{12}s_{23}s_{13}e^{i\delta} & s_{23}c_{13} \\
    s_{12}s_{23}-c_{12}c_{23}s_{13}e^{i\delta}
    & -c_{12}s_{23}-s_{12}c_{23}s_{13}e^{i\delta} & c_{23}c_{13}
  \end{pmatrix}
  \begin{pmatrix}
    e^{i\rho} & & \\
    & e^{i\sigma} & \\
    & & 1
  \end{pmatrix},
\end{equation}
where $c_{ij}\equiv\cos\theta_{ij}$, $s_{ij}\equiv\sin\theta_{ij}$, 
and $\delta$ is the CP-violating Dirac phase. The 
parameters $\rho$ and $\sigma$ are the Majorana phases, which are
irrelevant to neutrino flavor oscillations. In general, 
a $3\times3$ unitary matrix contains 3 rotation angles and 6 complex
phases, and a redefinition of lepton fields can remove 3 complex 
phases. Therefore the above parametrization covers the most general 
mixing matrix in the case that neutrinos are Majorana particles.

The information on the first and second generation neutrinos is
extracted, e.g.\ from the solar neutrino experiments~\cite{solar}
and the KamLAND reactor anti-neutrino experiment~\cite{KamLAND}. The
Super-Kamiokande experiment~\cite{SKatm}, the K2K long-baseline
experiment~\cite{K2K} and the CHOOZ reactor experiment~\cite{CHOOZ}
probe the energy region $\sim 10^{-(1-2)}$~eV and bring out the
information including the third generation neutrino. Recent global
analyses taking into account three-generation effects
suggest~\cite{analyses}
\begin{eqnarray}
\Delta m_{21}^2 &=& 7.1-8.9\times 10^{-5} \;\,{\rm eV}^2, \nonumber \\
|\Delta m_{31}^2| &=& 1.4-3.3\times 10^{-3} \;\,{\rm eV}^2, \nonumber \\
\sin^2\theta_{12} &=& 0.23-0.38, \nonumber \\
\sin^2\theta_{23} &=& 0.34-0.68, \nonumber \\
\sin^2\theta_{13} &<& 0.051,
\label{Nexp}
\end{eqnarray}
at the $3\sigma$ level. One interesting observation is that the best
fit value of the atmospheric angle $\theta_{23}$ is the maximal 
one ($45^\circ$), while the maximal mixture of the solar neutrinos is
disfavored more than the $5.6\sigma$ level~\cite{analyses}. As will be
seen in later sections, the significant deviation of the solar 
angle $\theta_{12}$ from the maximal one plays an important role to
search for the minimal forms of neutrino mass textures.

\section{Charged-fermion mass matrices}
\label{sec:minMc}
Before proceeding to the analysis of neutrino sector, we discuss the
charged lepton mass matrix in unified theory based on the minimal
quark mass matrices previously found. It turns out below that the
minimal form of charged-fermion mass matrices is uniquely determined
if one takes into account the discrepancy in mass eigenvalues between
down-type quarks and charged leptons. In this section, after briefly
reviewing the minimal forms of $M_u$ and $M_d$, we examine a way to
understand such discrepancy with group-theoretical factors in unified
theory.

\subsection{The minimal mass matrices for quarks}
In Ref.~\cite{asym}, we systematically investigated the minimality of
quark mass matrices $M_u$ and $M_d$, including the case that $M_d$ is
non-symmetric in the generation space. Here the 
term ``minimality'' means that the mass matrices successfully describe
the experimental data with as few numbers of non-vanishing elements as
possible. In other words, we have searched for the maximal number of
vanishing matrix elements such that they are consistent with the
observations. It is noted that the number of zero elements is that of
independently-vanishing elements in each matrix. For example, for a
symmetric matrix, ``1 zero'' implies that a diagonal element or a pair
of off-diagonal elements in symmetric positions takes a negligibly
small value compared to the others.
\begin{table}[ht]
\renewcommand{\arraystretch}{1.05}
\centering
\begin{tabular}{c|cc} \hline\hline
Solution & $M_u$ & $M_d$ \\ \hline
1 & ~~$\begin{pmatrix}
  a & & \\
  & & b \\
  & b & c
\end{pmatrix}$~~ &
~~$\begin{pmatrix}
  d& e& \\
  & & h \\
  & g& f
\end{pmatrix}$~~ \\ \hline
2 & $\begin{pmatrix}
  & a & \\
  a & & b \\
  & b & c
\end{pmatrix}$ &
$\begin{pmatrix}
  d& e & \\
  & & h \\
  & g & f
\end{pmatrix}$ \\ \hline
3 & $\begin{pmatrix}
  a & & \\
  & & b \\
  & b & c
\end{pmatrix}$ &
$\begin{pmatrix}
  & e & \\
  d & & h \\
  & g & f
\end{pmatrix}$ \\ \hline
4 & $\begin{pmatrix}
  & a & \\
  a & & b \\
  & b & c
\end{pmatrix}$ &
$\begin{pmatrix}
  & e & \\
  d & & h \\
  & g & f
\end{pmatrix}$ \\ \hline
5 & $\begin{pmatrix}
  a & & \\
  & & b \\
  & b & c
\end{pmatrix}$ &
$\begin{pmatrix}
  & e & \\
  & & h \\
  d & g & f
\end{pmatrix}$ \\ \hline
6 & $\begin{pmatrix}
  & a & \\
  a & & b \\
  & b & c
\end{pmatrix}$ &
$\begin{pmatrix}
  & e & \\
  & & h \\
  d & g & f
\end{pmatrix}$ \\ \hline\hline
\end{tabular}
\caption{The 6 patterns of successful minimal mass textures.}
\label{tab:MuMd}
\end{table}

Through the exhaustive analysis based on such minimality principle, we
have found that the 6 patterns of up and down quark matrices explain
all the observations in the quark sector. Table~\ref{tab:MuMd} is the
list of these 6 texture patterns. Each set of the minimal mass textures 
has 7 ($=3+4$) zero elements [i.e.\ 8 ($=3+5$) independent elements]
in the up and down quark mass matrices, respectively. Furthermore it
may be favorable that they are capable of leading to large leptonic 2-3
mixing (the atmospheric angle) via $SU(5)$ relation. We have also found
that there exist no solutions which explain large leptonic 1-2 mixing
(the solar angle) at this level of minimality. Therefore 
the 6 patterns in Table~\ref{tab:MuMd} should be fitted to the
experimental results for leptons by being combined with a relevant
form of neutrino mass matrices. When a large atmospheric angle
originates from the charged lepton sector, a more economical form of
neutrino mass matrices is expected to be found compared to the
analysis where the charged lepton mass matrix is flavor diagonal.

It is relevant here to comment on the existence of physically
equivalent mass textures. They are obtained by permuting rows and/or
columns of the mass matrices in Table~\ref{tab:MuMd}, i.e.\ by
exchanging generation indices 
of $u_L$, $u_R$, $d_L$ and/or $d_R$. There are two types of operations
which change the positions of zero matrix elements while preserving
physical consequences for mass eigenvalues and mixing angles: (i) the
exchanges of generation indices of $d_R$, and (ii) the exchanges of
the same generation indices of $u_L$ and $d_L$ (and also $u_R$ for a
symmetric $M_u$). Mass textures obtained by these operations are also
viable in the sense that they successfully describe the quark-sector
observables. Note that, if the quark mass matrices are embedded into
unified theory, a permutation of columns of $M_d$ (i.e.\ of $d_R$)
might affect physical quantity in the lepton sector since a
right-handed down-type quark and a left-handed charged lepton are
often unified in a single representation of unified gauge
group. However in the analysis including neutrino mass matrix, a
simultaneous exchange of generation indices 
of $d_R$, $e_L$, and $\nu_L$ preserves physical predictions also in the
lepton sector. Thus we conclude that it is sufficient to consider the
6 patterns presented in Table~\ref{tab:MuMd} and to examine all
flavor patterns of neutrino mass matrices, some of which are related
to each other by permutations of generation indices.

In addition to the above two operations (i) and (ii), there is an
approximate 2-3 exchange symmetry in $M_d$. In the parameter region
where $M_u$ and $M_d$ properly reproduce the data of the quark sector,
all $M_d$ in Table~\ref{tab:MuMd} lead to large 2-3 mixing of $d_R$ so 
that $V_{dR}$ becomes approximately invariant under the exchange of
second and third rows. As will be seen, this feature would lead to
some counterintuitive consequences in the analysis of neutrino mass
matrices. We will mention this point in later discussion.

We have also found in the previous work the other possibilities of
quark mass textures which have the same number of vanishing matrix 
elements (7 zeros) as the solutions in Table~\ref{tab:MuMd}:
\begin{eqnarray}
  M_u &=& \begin{pmatrix}
    & a & \\
    a & d & b \\
    & b & c
  \end{pmatrix}, \;\; 
  \begin{pmatrix}
    & & a \\
    & b & d \\
    a & d & c
  \end{pmatrix} \\[1mm]
  M_d &=& \begin{pmatrix}
    e & h & \\
    & f & \\
    & & g
  \end{pmatrix}, \;\; 
  \begin{pmatrix}
    e & & \\
    h & f & \\
    & & g
  \end{pmatrix}
  \label{otherMuMd}
\end{eqnarray}
All 4 ($=2\times2$) combinations of $M_u$ and $M_d$ well describe the
present experimental data of the quark sector. An important difference
between these 4 patterns and those in Table~\ref{tab:MuMd} is the
structure of down-quark mass matrix: the matrices 
in (\ref{otherMuMd}) do not have the 2-3 generation mixture and thus
no significant contribution to the atmospheric angle arises when the
matrices are embedded into $SU(5)$ framework. We checked that this
fact actually needs non-minimal, complex forms of neutrino mass
matrices. In this paper, therefore, we concentrate on the mass
textures shown in Table~\ref{tab:MuMd} in the viewpoint of minimality.

\subsection{The minimal mass matrix for charged leptons}
In unified theory, it may be natural to assume matter unification
which often leads to the same mass eigenvalues for down-type quarks
and charged leptons. However when the observed values of these masses
are extrapolated up to high-energy scale by renormalization group
evolution, the following relations are found to be satisfied at the
unification scale:
\begin{equation}
  m_d \,\sim\, 3m_e, \qquad 3m_s \,\sim\, m_\mu, \qquad 
  m_b \,\sim\, m_\tau.
  \label{dl_ratio}
\end{equation}
To reproduce these mass relations, especially for the first and second
generations, it is needed to introduce some unified symmetry breaking
which splits the properties of quarks and leptons. A typical example
of such breaking effects is provided by group-theoretical 
factors~\cite{GJ}, which originate in Yukawa interactions involving
additional higher-dimensional Higgs fields. In this subsection, we
examine the embedding of the minimal quark mass textures into unified
theory, taking account of the effect of symmetry-breaking
factors. There might be other possibilities for realizing 
(\ref{dl_ratio}) such as contributions from higher-dimensional
operators, which bring the Yukawa sector additional free parameters to
be tuned. These operators are also suppressed by the ratio between the
unification scale and the gravitational scale. In connection with the
minimality principle, we focus on the effect of group-theoretical 
factors which do not necessarily require additional parameters
describing masses and mixing angles.

Let us first closely examine the minimal down-quark mass 
matrix $M_d$. There are 3 types of $M_d$ in 
Table~\ref{tab:MuMd}: $M_d^{\mbox{\tiny(1)}}$ for 
Solutions~1,2, $M_d^{\mbox{\tiny(2)}}$ for Solutions~3,4, 
and $M_d^{\mbox{\tiny(3)}}$ for Solutions~5,6. The mass hierarchy and
mixing angles are properly reproduced with the following 
parametrizations:
\begin{eqnarray}
  \frac{M_d^{\mbox{\tiny(1)}}}{m_b} &=&
  \begin{pmatrix}
    \bar d\lambda^4 & \bar e\lambda^3 & \\
    & & \bar h\lambda^2 \\
    & \bar g & \bar f \\
  \end{pmatrix}, \\[1mm]
  \frac{M_d^{\mbox{\tiny (2)}}}{m_b} &=& \begin{pmatrix}
    & \bar e\lambda^3 & \\
    \bar d\lambda^3 & & \bar h\lambda^2 \\
    & \bar g & \bar f \\
  \end{pmatrix}, \\[1mm]
  \frac{M_d^{\mbox{\tiny(3)}}}{m_b} &=&
  \begin{pmatrix}
    & \bar e\lambda^3 & \\
    & & \bar h\lambda^2 \\
    \bar d\lambda & \bar g & \bar{f} \\
  \end{pmatrix},
\end{eqnarray}
where $\lambda$ is a small parameter of the order of the Cabibbo 
angle ($\lambda\sim0.2$), and $\bar d$, $\bar e$, $\bar f$, $\bar g$,
$\bar h$ are dimensionless $\mathcal{O}(1)$ coefficients. It is found
in Ref.~\cite{asym} that the quark masses and mixing angles can be
fitted by adjusting these coefficients. For any of the above forms 
of $M_d$, the mass eigenvalues for the second and third generations
satisfy
\begin{equation}
  \frac{m_s}{m_b} \;\simeq \; \bar g\bar h\lambda^2
  \label{msmb}
\end{equation}
at the leading order of the expansion parameter $\lambda$. On the
other hand, the mass eigenvalue of the first generation has rather
different expressions:
\begin{eqnarray}
  \frac{m_d}{m_b} \;\simeq\;\left\{
  \begin{array}{ccl}
    \bar d\lambda^4 && \textrm{for } M_d^{\mbox{\tiny(1)}} \\[2mm]
    \frac{\bar d\bar e\bar f}{\bar g\bar h}\lambda^4 &&
    \textrm{for } M_d^{\mbox{\tiny(2)}} \\[2.5mm]
    \frac{\bar d\bar e}{\bar g}\lambda^4 && 
    \textrm{for } M_d^{\mbox{\tiny(3)}}
  \end{array} \right.
  \label{mdmb}
\end{eqnarray}

In unified theory framework, the charged lepton mass 
matrix $M_e$ naively has the same form as $M_d$, which does not lead
to the correct mass parameters in low-energy regime. To resolve it, we
introduce the group-theoretical factor~\cite{GJ} assuming Higgs fields
in higher-dimensional representations, e.g.~45 plet.~of $SU(5)$. That
induces relative factors ``$-3$'' in $M_e$ compared to the
corresponding matrix elements in $M_d$.

It is first found from (\ref{dl_ratio}) and (\ref{msmb}) that the 
element $\bar h$ should be replaced with $-3\bar h$ in $M_e$ to bridge
the gap between the strange quark and muon masses. Notice that the
alternative choice, i.e.~the factor $-3$ for $\bar g$, may disturb the
prediction $m_b\simeq m_\tau$, and we do not consider such a
possibility in this paper. As for the first-generation mass
eigenvalues, the relation (\ref{dl_ratio}) tells us that the electron
mass is suppressed by the factor 3 compared to the down-quark mass. It
is seen from (\ref{mdmb}) that this suppression can be realized only 
for $M_d^{\mbox{\tiny(2)}}$ with the 
replacement $\bar h\to-3\bar h$. The 
replacement $\bar g\to-3\bar g\,$ for $M_d^{\mbox{\tiny(2),(3)}}$ also
gives the same suppression effect for the electron mass, but in this
case, the bottom-tau unification does not follow. We thus find that
to introduce a group-theoretical factor $-3$ for the 
element $\bar h$ in Solutions 3,4 is the unique choice to properly
reproduce the observed values of charged lepton masses at low
energy. It may be interesting to notice that the resolution is
economical in that a single attachment factor is sufficient to
accommodate mass eigenvalues to the experimental relations
(\ref{dl_ratio}) and no other modification is allowed for the minimal
mass matrices.

\begin{figure}[t]
\centering
\includegraphics[scale=1.5]{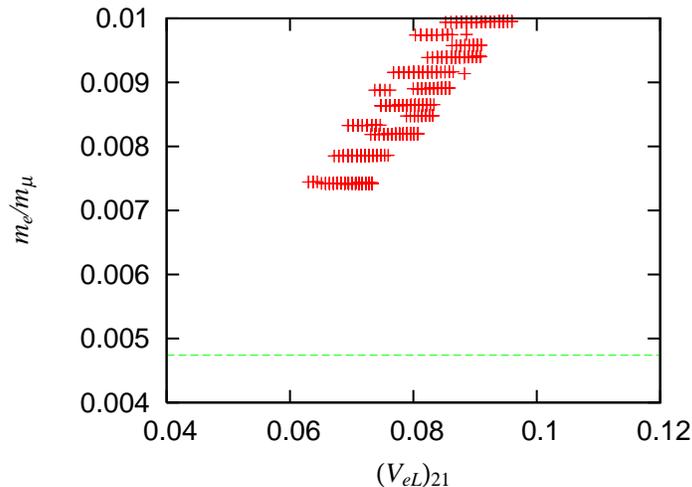}
\caption{The predictions of the mass ratio $m_e/m_\mu$ and mixing
matrix element $(V_{eL})_{21}$ for Solution 3. The horizontal dashed
line shows the experimental value of the mass ratio.\bigskip}
\label{fig:sol3}
\end{figure}

\begin{figure}[t]
\centering
\includegraphics[scale=1.5]{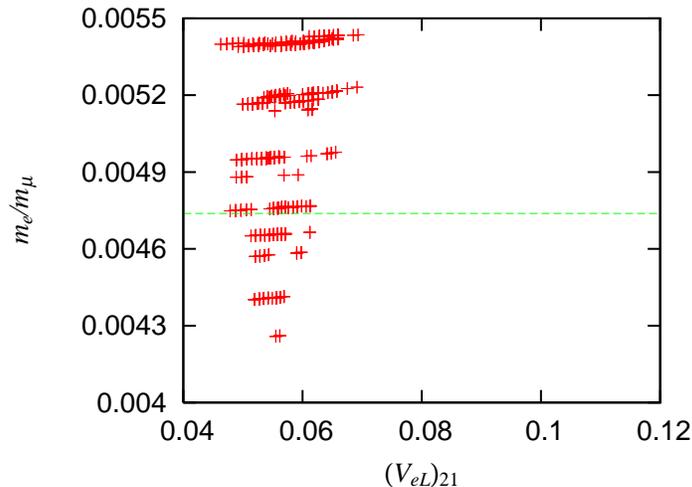}
\caption{The same as Fig.~\ref{fig:sol3} for Solution 4.}
\label{fig:sol4}
\end{figure}

We have shown that only Solutions 3 and 4 embedded in unified theory
are consistent with the observed values of down-type quark and
charged lepton masses. Let us see whether these solutions have
parameter space where all other experimental constraints including the
flavor mixing angles are satisfied. In 
Figs.~\ref{fig:sol3} and \ref{fig:sol4}, we show the parameter regions
in which the quark mass data (\ref{massexp}) and the experimental
results (\ref{CKMexp})$-$(\ref{sin2b}) are explained. The horizontal
axis shows the 1-2 generation mixture of left-handed charged leptons
and the vertical one the mass ratio $m_e/m_\mu$. The horizontal
dashed lines denote the observed value of $m_e/m_\mu$ at the
electroweak scale. It is found from these figures that 
Solution 4 accounts for the observed data in the quark sector as well
as the charged lepton mass relations, but Solution 3 does not. We
numerically checked that Solution 4 does explain all the quark-sector
constraints listed in Section~\ref{sec:exp} and also charged lepton
mass eigenvalues.

There are two comments on the above analysis. First,
Figs.~\ref{fig:sol3} and \ref{fig:sol4} are drawn by using the
experimental data at the electroweak scale. Since we are working in
unified theory framework, the mass matrices are supposed to be given
at the unification scale and should be compared to the data at that
scale. However the conclusion obtained from 
Figs.~\ref{fig:sol3} and \ref{fig:sol4} are stable against
renormalization group effects. This is because radiative corrections 
to $(V_{eL})_{21}$ and $m_e/m_\mu$ do not involve any large coupling
constants, and these quantities are almost energy-scale independent.
The second is the prediction for the mixing matrix 
element $(V_{eL})_{21}$, i.e.\ the 1-2 generation mixing of
left-handed charged leptons. It is found from Fig.~\ref{fig:sol4}
that, for the successful Solution 4, it takes the value $\sim 0.07$ at
most, which is not so large to solely provide the observed value of
the solar angle. This fact will play an important role in classifying
neutrino mass matrices in later sections.

\bigskip

In summary, we have found that only Solution 4 in Table~\ref{tab:MuMd}
can be accommodated to the charged lepton masses without contradicting
the observations in the quark sector (\ref{massexp})$-$(\ref{sin2b}),
if a group-theoretical factor is properly applied. The minimal mass
matrices for charged fermions are thus given by
\begin{equation}
  M_u \,=\, \begin{pmatrix}
    & a & \\
    a & & b \\
    & b & c
  \end{pmatrix},\quad\;
  M_d \,=\, \begin{pmatrix}
    & e & \\
    d & & h \\
    & g & f
  \end{pmatrix},\quad\;
  M_e \,=\, \begin{pmatrix}
    & e & \\
    d & & \!\!-3h \\
    & g & f
  \end{pmatrix}.
  \label{uniqueMc}
\end{equation}
An interesting prediction of this combination 
of $M_u$, $M_d$ and $M_e$ is a large 2-3 mixing angle in the
charged lepton mixing matrix $V_{eL}$. In the parameter region where
the experimental data of the quark sector is properly 
reproduced, $V_{eL}$ is approximately invariant under the exchange of
the second and third rows. Therefore a neutrino mass texture which is
obtained by exchanging the generation labels of $L_2$ and $L_3$ 
predicts almost the same physics as the original one.

\section{Majorana mass matrices for left-handed neutrinos}
In this section, we study the matrix form of left-handed neutrino
Majorana masses. The Majorana mass matrix $M_L$ defined in (\ref{ML})
are derived in various ways, for example, from the higher-dimensional
operator $HHL^{\rm T}_iC L_j$, the expectation value of electroweak
triplet scalar field, etc. Keeping in mind such dynamical origins, we
explore the minimality of effective mass matrix $M_L$, combining it
with the successful minimal textures of charged fermions
(\ref{uniqueMc}). Here we use the minimality in the same meaning as in
Section~\ref{sec:minMc}, that is, the maximal number of zero matrix
elements which is consistent to all the experimental results for
quarks and leptons.

\subsection{The would-be minimal forms}
\label{sec:minML}
Let us first examine the minimality of $M_L$ by taking account of the
information about neutrino mass eigenvalues only. Since only two
mass-squared differences are measured in neutrino oscillation 
experiments, one of three-generation neutrinos can be exactly
massless. Thus we expect that the minimal matrix $M_L$ would have one
zero and two nonzero mass eigenvalues to reproduce the 
data (\ref{Nexp}). This fact is a crucial difference from the texture
analysis of quark mass matrices. Next, consider the neutrino
generation mixing described by the matrix $V_\nu$ which 
diagonalizes $M_L$. As for the generation mixing, 
the solution (\ref{uniqueMc}) provides a large 2-3 mixing for the
lepton sector. Furthermore we have found that the 1-2 mixing of
charged leptons is not so large enough to explain the observed solar
neutrino oscillation. Therefore the neutrino mixing 
matrix $V_\nu$ must involve a sizable mixture between the first and
second generations.

The above discussion about neutrino masses and mixing restricts the
candidates for the minimal matrix to the following forms:
\begin{equation}
  M_L \,=\, \begin{pmatrix}
    & l & ~\\
    l & m & \\
    & & 
  \end{pmatrix}, \;\;
  \begin{pmatrix}
    m & l & ~\\
    l & & \\
    & &
  \end{pmatrix}.
  \label{minML}
\end{equation}
Note that these two types of matrices give different physical
predictions since the basis of left-handed leptons is already fixed 
by $M_e$. In addition to these matrices, there are two additional
candidates for the minimal mass texture which are obtained by
exchanging the second and third generation indices 
in (\ref{minML}). At first glance, it seems that those additional
candidates provide only 1-3 generation mixing and do not work for
viable solutions. However they should be included in the list since,
as we mentioned before, the charged lepton mixing matrix $V_{eL}$ for
Solution 4 is approximately invariant under the permutation of the 
second and third rows. Therefore a neutrino mass matrix $M_L$ always
has such a doubling freedom while keeping physical consequences. In
the following, we do not explicitly write down doubled candidates,
though they are included in our analysis.

\begin{figure}[t]
\centering
\includegraphics[scale=1.5]{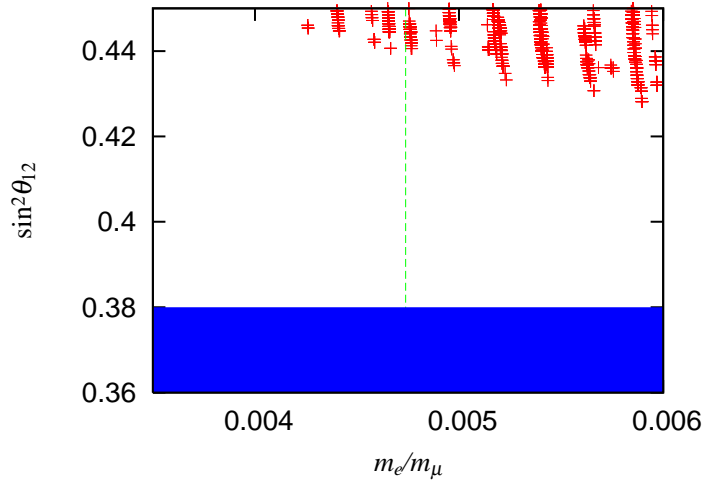}
\caption{A typical prediction of the matrix forms (\ref{uniqueMc}) and
(\ref{minML}). The horizontal and vertical axes mean the mass 
ratio $m_e/m_\mu$ and the solar angle, respectively. The filled (blue)
region is experimentally allowed for $\theta_{12}$ at 
the $3\sigma$ level. The vertical dashed-line shows the observed value
of the mass ratio.\bigskip}
\label{fig:min}
\end{figure}

The matrices (\ref{minML}) seem to satisfy the criterions for neutrino
masses and mixing, but it turns out that they are impractical if
combined with the minimal solution (\ref{uniqueMc}). For example, let
us see the first mass matrix in (\ref{minML}). (For the second matrix,
a similar discussion can also be applied.) \ It is first noticed that
the matrix predicts the spectrum that the third-generation neutrino is
massless and the other two neutrinos should have almost degenerate
masses. That is,
\begin{equation}
  \frac{m_1}{m_2} \,\simeq\, 
  1-\frac{1}{2}\frac{\Delta m_{21}^2}{\Delta m_{23}^2}, \qquad
  m_3 \,=\, 0.
\end{equation}
A non-vanishing off-diagonal element $(V_\nu)_{12}$ is expressed in
terms of the mass eigenvalues:
\begin{equation}
  (V_\nu)_{12} \,=\, \sqrt{\frac{m_1}{m_1 + m_2}}\,.
\end{equation}
Thus the 1-2 mixing angle in $V_{\rm MNS}$ is given by
\begin{equation}
  \sin\theta_{12} \,\simeq\, \frac{1}{\sqrt{2}}
  -\frac{1}{\sqrt{2}}(V_{eL})_{21},
  \label{min_t12}
\end{equation}
up to tiny corrections of order ${\mathcal O}(10^{-3})$. We have
assumed $(V_{eL})_{11}\simeq1$ and taken phase degrees of freedom in
Yukawa couplings so that the last term in (\ref{min_t12}) is
destructive to account for $\theta_{12}<45^\circ$. The experimental
value of $\theta_{12}$ (\ref{Nexp}) gives a bound on the magnitude of
charged lepton mixing:
\begin{equation}
  0.13 \;<\; (V_{eL})_{21} \;<\; 0.32
  \label{VeL_bound}
\end{equation}
at the $3\sigma$ level. Therefore a crucial point required for the
minimal mass matrices is whether a sizable contribution 
to $(V_{eL})_{21}$ arises or not. We find from (\ref{uniqueMc}) an
approximate analytic expression for charged lepton mixing:
\begin{equation}
  (V_{eL})_{21} \;<\; \frac{1}{3}
  \bigg(V_{us}+\sqrt{\frac{m_u}{m_c}}\bigg) \;\sim\,0.09,
  \label{VeL21}
\end{equation}
which is not consistent to the bound (\ref{VeL_bound}) (and see also
the numerical result, Fig.~\ref{fig:sol4}). The 
inequality (\ref{VeL21}) implies that the group-theoretical factor 
in $M_e$ prevents the mixing $(V_{eL})_{21}$ from being large enough
to fulfill the required value (\ref{VeL_bound}). In
Fig.~\ref{fig:min}, we present a typical numerical result of the
combination of (\ref{uniqueMc}) and the first matrix 
in (\ref{minML}). All the other experimental constraints are satisfied
in this figure. It is found that the 1-2 mixing of charged leptons
cannot be so large, which confirms the above analytic result. In the
end, the left-handed Majorana mass matrix $M_L$ (\ref{minML}) are
experimentally disfavored at more than $3\sigma$ level.

\subsection{The minimal mass matrix for Majorana neutrinos}
\label{sec:nxtminML}
We have shown that two would-be minimal patterns of mass matrices
(\ref{minML}) are not suitable to explain the solar neutrino
oscillation. Let us consider the next level of minimality for the
Majorana mass matrix $M_L$ which has 3 vanishing matrix
elements. There are 20 patterns of symmetric matrices 
with 3 zeros. By a detailed analysis of these 20 matrices, the
following 8 patterns are found to be successful:
\begin{equation}
  M_L \;=\;
  \begin{pmatrix}
    & l & \\
    l & m & \\
    & & n
  \end{pmatrix}, \;
  \begin{pmatrix}
    & l & \\
    l & & m \\
    & m & n
  \end{pmatrix}, \;
  \begin{pmatrix}
    l & m & \\
    m & n & \\
    & & ~
  \end{pmatrix}, \;
  \begin{pmatrix}
    & & l \\
    & & m \\
    l & m & n
  \end{pmatrix},
  \label{nxtminML}
\end{equation}
and four matrices obtained by exchanging the second and
third-generation indices in (\ref{nxtminML}). That is, the 8
matrices are classified into 4 types which are physically
inequivalent. Any combination of (\ref{uniqueMc}) and one of these
matrices is in accordance with the observed data of quarks and leptons
(\ref{massexp})$-$(\ref{sin2b}) and (\ref{Nexp}). In the following, we
discuss in order the phenomenological implications of the minimal
forms of $M_L$ (\ref{nxtminML}). They have rather different
predictions for physical quantities and can be distinguished. The
predictions are summarized in the end of this section (see
Table~\ref{tab:predict}).
\begin{enumerate}
\item
\begin{equation}
\begin{pmatrix}
  & l & \\
  l & m & \\
  & & n
\end{pmatrix}\qquad
\left[\begin{pmatrix}
  & & l \\
  & n & \\
  l & & m
\end{pmatrix}\right]
\label{nxtminML1}
\end{equation}
The latter matrix (in the bracket) is obtained by exchanging the
second and third labels of the former and has similar physical
implications due to the approximate invariance of $V_{eL}$ as
mentioned before. It is first noticed that the 
matrix (\ref{nxtminML1}) is able to provide only 1-2 neutrino
mixing. Thus the physical mixing angle $\theta_{13}$ is given by
\begin{equation}
  \sin\theta_{13} \;=\; |(V_{eL})_{31}| \;\simeq\; 0.025-0.055
\end{equation}
for the minimal solution (\ref{uniqueMc}). The numerical uncertainty
mainly comes from those of the down and strange quark masses. As for
the neutrino 1-2 mixing, we find an approximate relation:
\begin{equation}
  \sin\theta_{12} \;\simeq\; |(V_\nu)_{12}| \;=\,
  \sqrt{\frac{m_1}{m_1 + m_2}}\,,
  \label{nxtmin_t12}
\end{equation}
where we have neglected $\mathcal{O}(10^{-2})$ corrections from the
charged lepton side.

~~~~In general, there are 3 types of neutrino mass spectrum allowed by
the current experimental data: normal hierarchy ($m_1<m_2\ll m_3$),
inverted hierarchy ($m_3\ll m_1<m_2$), and degenerate 
one ($m_1\simeq m_2\simeq m_3$). For the mass matrix
(\ref{nxtminML1}), the eigenvalues with normal hierarchy are
expressed in terms of the observables in oscillation experiments:
\begin{eqnarray}
  m_1 &\simeq& 
  \sqrt\frac{\Delta m^2_{21}\tan^4\theta_{12}}{1-\tan^4\theta_{12}},
  \nonumber \\
  m_2 &\simeq& 
  \sqrt{\frac{\Delta m^2_{21}}{1-\tan^4\theta_{12}}}, \nonumber \\
  m_3 &\simeq& \sqrt{\Delta m^2_{32}
    +\frac{\Delta m^2_{21}}{1-\tan^4\theta_{12}}}.
  \label{normal}
\end{eqnarray}
On the other hand, it is easily found from (\ref{nxtmin_t12}) that the
inverted hierarchy and degenerate spectrum lead to a nearly maximal 
mixing angle $\sin\theta_{12}\sim1/\sqrt{2}$, and therefore cannot be
predicted by the mass texture (\ref{nxtminML1}). Finally, the
effective Majorana mass for neutrino-less double beta decay is given
by $|\langle m_{ee}\rangle|\simeq
\sqrt{\Delta m_{21}^2\sin2\theta_{12}\tan2\theta_{12}}\,
|(V_{eL})_{21}|$, and is found to be small.
\item
\begin{equation}
\begin{pmatrix}
  & l & \\
  l & & m \\
  & m & n
\end{pmatrix}\qquad
\left[\begin{pmatrix}
  & & l \\
  & n & m \\
  l & m &
\end{pmatrix}\right]
\end{equation}
The latter in the bracket is the label-exchanged matrix from the
former. Diagonalizing the mass matrix, we find the off-diagonal
elements $(V_\nu)_{12}$ and $(V_\nu)_{13}$ are given by
\begin{eqnarray}
  (V_\nu)_{12} &=& 
  \sqrt{\frac{m_1(m_1^2-m_3^2)}{(m_1+m_2)(m_1^2-m_1m_2+m_2^2-m_3^2)}},\\
  (V_\nu)_{13} &=& 
  \sqrt{\frac{m_1m_2(m_1-m_2)}{(m_1-m_3)(m_2+m_3)(m_1-m_2+m_3)}}.
\end{eqnarray}
If the first-two generations are almost 
degenerate ($m_1\simeq m_2$), i.e.\ for the inverted hierarchy or
degenerate spectrum, the above relation 
indicates $(V_\nu)_{12}\simeq 1/\sqrt{2}$. That is unfavorable for the
solar neutrino data since the electron has only slight mixture to the 
other generations in the minimal solution (\ref{uniqueMc}). On the
other hand, for the normal hierarchy ($m_{1,2}\ll m_3$), the 
formula of $(V_\nu)_{12}$ is reduced to the 
relation (\ref{nxtmin_t12}) at the leading order of $m_{1,2}/m_3$, and
also the mass spectrum (\ref{normal}) is reproduced up to small
corrections. In addition, the neutrino 1-3 mixing angle becomes at the
leading order
\begin{equation}
  (V_\nu)_{13} \;\simeq\; \frac{m_2}{m_3}\sqrt{\frac{m_1}{m_3}}
  \;\simeq\; \frac{\tan\theta_{12}}{(1-\tan^4\theta_{12})^{3/4}}
  \bigg(\frac{\Delta m_{21}^2}{\Delta m_{32}^2}\bigg)^{\!\frac{3}{4}}
  \;\simeq\; 0.04-0.14. 
\end{equation}
Thus $\sin\theta_{13}$ is of this order when there is no accidental
parameter cancellation between the charged lepton and neutrino
sectors, and would be detectable in future neutrino experiments.
\item
\begin{equation}
\begin{pmatrix}
  l & m & \\
  m & n & \\
  & & ~
\end{pmatrix}\qquad
\left[\begin{pmatrix}
  l & & m \\
  & & \\
  m & & n
\end{pmatrix}\right]
\end{equation}
An important feature of this mass texture is a vanishing determinant,
which predicts the inverted hierarchy with a massless third-generation
neutrino. As for the mixing angles, no 1-3 neutrino mixing follows and
the physical lepton 1-3 mixing is therefore given by
\begin{equation}
  \sin\theta_{13} \;=\; |(V_{eL})_{31}| \;\simeq\; 0.025-0.055  
\end{equation}
for the minimal solution (\ref{uniqueMc}). Since the matrix has 3 free
parameters for the first-two generations, there are no predictive
relations, in particular, for the lepton 1-2 mixing angle. It is also
interesting to notice that the 1-1 element of the matrix is nonzero,
which might give a sizable rato of neutrino-less double beta 
decay ($0\nu\beta\beta$). With a massless third-generation neutrino,
the averaged mass parameter responsible for $0\nu\beta\beta$ is
evaluated as
\begin{equation}
  \langle m_{ee}\rangle \,\simeq\; m_1\cos^2\theta_{12}e^{-2i\rho}
  +m_2\sin^2\theta_{12}e^{-2i\sigma},
\end{equation}
where $\rho$ and $\sigma$ are the Majorana phases of neutrinos 
in $V_{\rm MNS}$, and we have used $\sin\theta_{13}\ll 1$. In the
parameter region where the experimental data (\ref{Nexp}) is
explained, the averaged neutrino mass takes the 
value $|\langle m_{ee}\rangle|\simeq 0.0084-0.058$~eV\@. The planned
improvements in the sensitivity to such a small mass parameter are
expected to reach $|\langle m_{ee}\rangle|\sim
\mathcal{O}(10^{-2})$~eV~\cite{2beta}, and this type of minimal matrix
will be testable in near future.
\item
\begin{equation}
\begin{pmatrix}
  & & l \\
  & & m \\
  l & m & n
\end{pmatrix}\qquad
\left[\begin{pmatrix}
  & l & \\
  l & n & m \\
  & m &
\end{pmatrix}\right]
\label{nxtminML4}
\end{equation}
This matrix also has a vanishing determinant. We find that it predicts
the spectrum with normal hierarchy and a massless first-generation
neutrino. The inverted mass hierarchy is not allowed 
since $(V_\nu)_{12}\simeq 1/\sqrt{2}$ in the parameter region where
the CHOOZ bound in (\ref{Nexp}) is satisfied, which leads to a too
large solar angle. Another point of the matrix (\ref{nxtminML4}) is
that it induces a relatively large value of lepton 1-3 mixing
angle. For the normal hierarchy spectrum, one obtains a relation in
the neutrino sector:
\begin{equation}
  (V_\nu)_{13} \,=\, (V_\nu)_{12}\sqrt{\frac{m_2}{m_3}},
\end{equation}
which gives the following prediction among the observables:
\begin{equation}
  \sin^2\theta_{13} \,\simeq\, \sin^2\theta_{12}
  \sqrt{\frac{\Delta m^2_{21}}{\Delta m^2_{32}}},
\end{equation}
up to small corrections from the charged lepton sector. Substituting
the experimental data (\ref{Nexp}) for the right-handed side, we
roughly obtain $\sin\theta_{13}\simeq 0.18-0.32$, which is marginal to
the current experimental upper bound. The effective Majorana mass is
predicted to be small: $|\langle m_{ee}\rangle|\simeq 
2(\Delta m_{21}^2\Delta m_{32}^2)^{\frac{1}{4}}
|(V_\nu)_{12}(V_{eL})_{31}|<1.8\times 10^{-3}$ eV, if taken account of
the experimental bound on $\theta_{13}$.
\end{enumerate}
\begin{table}[t]
\renewcommand{\arraystretch}{1.05}
\centering
{\small%
\begin{tabular}{c|c|c|c|c|c}\hline\hline
Solution & $m_1$ (eV) & $m_2$ (eV) & $m_3$ (eV) & $\sin\theta_{13}$ & 
$|\langle m_{ee}\rangle|\,$ (eV) \\ \hline
$\!\begin{pmatrix} & l& \\ l & m & \\ & & n \end{pmatrix}\!$ & 
$\!0.0026-0.0075\!$ & $0.0088-0.012$ & $0.038-0.059$ &
$\!0.025-0.055\!$ & $\!0.0003-0.0011\!$ \\ \hline
$\!\begin{pmatrix} & l & \\ l & & m \\ & m & n \end{pmatrix}\!$ & 
$0.0027-0.010$ & $0.0089-0.014$ & $0.040-0.060$ & $0-0.15$ & 
$\!0.0002-0.0014\!$ \\ \hline
$\!\begin{pmatrix} l & m & \\ m & n & \\ & & ~\end{pmatrix}\!$ &
$0.035-0.058$ & $0.036-0.058$ & $0$ & $\!0.025-0.055\!$ & 
$0.0084-0.058$ \\ \hline
$\!\begin{pmatrix} & & l \\ & & m \\ l & m & n \end{pmatrix}\!$ &
$0$ & $\!0.0084-0.0094\!$ & $0.037-0.059$ & $0.13-0.37$ & 
$\!0.0002-0.0018\!$ \\ \hline\hline
\end{tabular}}
\caption{The predictions of the minimal Majorana mass 
matrices $M_L$ with the minimal solution (\ref{uniqueMc}). A matrix
obtained by exchanging the second and third labels in each solution is
also available which has almost the same physical predictions as the
original one. In the first line, $m_{1,2,3}$ are the mass eigenvalues,
and $\langle m_{ee}\rangle$ is the averaged mass parameter for
neutrino-less double beta decay.\bigskip}
\label{tab:predict}
\end{table}

\section{The minimal mass matrices in the seesaw scenario}
\label{sec:minMnMR}
In the previous section, we have investigated possible forms of the
Majorana mass matrix of left-handed neutrinos without referring to
underlying mechanisms which realize tiny neutrino masses. One of the
most attractive schemes for light neutrinos is the seesaw
mechanism~\cite{seesaw} to introduce heavy right-handed neutrinos and
consider the following mass terms:
\begin{equation}
  -{\cal L} \,=\, \overline{\nu_R}{}_i(Y_\nu)_{ij} L_jH
  +\frac{1}{2}\nu_R{}_i^{\rm T}(M_R)_{ij}C \nu_R{}_j +\textrm{h.c.},
\end{equation}
where $\nu_{R_i}$ denote right-handed neutrinos which are singlets
under the standard model gauge group, $Y_\nu$ is the neutrino Yukawa
matrix, and $M_R$ the Majorana mass matrix of right-handed
neutrinos. If mass eigenvalues of $M_R$ are sufficiently larger than
the electroweak scale, one obtains effective Majorana mass matrix of
left-handed neutrinos at low-energy regime:
\begin{equation}
  M_L \,=\, -M_\nu^{\rm T} M_R^{-1} M_\nu,
  \label{seesawML}
\end{equation}
where $M_\nu$ is the neutrino Dirac mass 
matrix: $M_\nu=Y_\nu\langle H\rangle$.

In this section, we examine the minimality of two mass
matrices $M_\nu$ and $M_R$, that is, the minimal (total) number of
zero matrix elements which are compatible with the experimental
data. Note that the minimality of $M_\nu$ and $M_R$ is different from
that of $M_L$: as will be seen, the minimal forms 
of $M_\nu$ and $M_R$ do not necessarily lead to those of $M_L$ found in
the previous section. Therefore these two analyses of minimality
should be independently performed. We assume that there exist no
sterile states in the light spectrum. That implies the Majorana mass
matrix $M_R$ has a non-vanishing determinant, which is a contrast to
the previous analysis where $M_L$ may have a vanishing determinant. It
is seen from the seesaw formula (\ref{seesawML}) that any exchange of
the generation indices of right-handed neutrinos does not change
physical consequences at low-energy regime. Further the label exchange
of left-handed neutrinos $\nu_{L_2}\leftrightarrow\nu_{L_3}$ also
gives additional viable textures because of the approximate 2-3
exchanging symmetry of $V_e$ as noted before. In the following, we do
not particularly mention physically-equivalent label-exchanged
matrices, though the results should always be interpreted as including
such additional solutions.

\medskip

It is first noticed that the minimal form of $M_R$ with a
non-vanishing determinant has 4 independent zeros. There are 3 types
of matrix forms with such symmetric 4 zeros:
\begin{equation}
  M_R \,=\,
  \begin{pmatrix}
    & r & \\
    r & & \\
    & & s
  \end{pmatrix}, \;
  \begin{pmatrix}
    & & r \\
    & s & \\
    r & &
  \end{pmatrix}, \;
  \begin{pmatrix}
    s & & \\
    & & r \\
    & r &
  \end{pmatrix}.
  \label{4zeroMR}
\end{equation}
They are related to each other by permuting generation labels. The
Dirac mass matrix $M_\nu$ must have at least two nonzero eigenvalues
to account for the neutrino oscillation data. Therefore $M_\nu$ with 7
vanishing entries are the simplest candidates for minimality. There
are 3 classes of such rank-two matrix $M_\nu$ with 7 zeros:
\begin{equation}
  M_\nu \,=\,
  \begin{pmatrix}
    p & & \\
    & q & \\
    & & ~ 
  \end{pmatrix}, \;
  \begin{pmatrix}
    ~ & & \\
    & p & \\
    & & q
  \end{pmatrix}, \;
  \begin{pmatrix}
    p & & \\
    & ~ & \\
    & & q
  \end{pmatrix}.
  \label{7zeroMn}
\end{equation}
All the other matrices obtained by exchanging rows and/or columns need
not to be included since the relabeling effect is already taken into 
account in $M_R$ (\ref{4zeroMR}). It is easily found that, via the
seesaw operation (\ref{seesawML}), any combination of 
4-zero $M_R$ (\ref{4zeroMR}) and 7-zero $M_\nu$ (\ref{7zeroMn}) gives
an effective Majorana mass matrix $M_L$ with only one mass squared
difference, and therefore is excluded by the oscillation experiment
results.

The next simplest candidate is the combination 
of $M_\nu$ and $M_R$ with 10 vanishing matrix elements in 
total: 6-zero $M_\nu\,+$ symmetric 4-zero $M_R$ and 
7-zero $M_\nu\,+$ symmetric 3-zero $M_R$. We examined all matrix
patterns whether they can totally fit the experimental data of quarks
and leptons, and found that the following mass textures are almost
successful: 
\begin{itemize}
\item $M_\nu$ with 6 zeros + $M_R$ with symmetric 4 zeros
\begin{eqnarray}
M_\nu &=&
\begin{pmatrix}
  p & & ~ \\
  & q & \\
  r & &
\end{pmatrix}, \;
\begin{pmatrix}
  & p & ~ \\
  q & & \\
  & r &
\end{pmatrix}, \;
\begin{pmatrix}
  & & ~ \\
  p & q& \\
  & r &
\end{pmatrix}, \;
\begin{pmatrix}
  & & ~ \\
  p & q & \\
  r & &
\end{pmatrix}, \\
M_R &=&
\begin{pmatrix}
  s & & \\
  & & t \\
  & t &
\end{pmatrix}.
\end{eqnarray}
\item $M_\nu$ with 7 zeros + $M_R$ with symmetric 3 zeros
\begin{eqnarray}
M_\nu &=&
\begin{pmatrix}
  p & & \\
  & q & \\
  & & ~
\end{pmatrix}, \\
M_R &=&
\begin{pmatrix}
  & r & \\
  r & s & \\
  & & t
\end{pmatrix}, \;
\begin{pmatrix}
  & r & \\
  r & & s \\
  & s & t
\end{pmatrix}, \;
\begin{pmatrix}
  r & s & \\
  s & & \\
  & & t
\end{pmatrix}, \;
\begin{pmatrix}
  & r & s \\
  r & & \\
  s & & t
\end{pmatrix}.
\end{eqnarray}
\end{itemize}
In addition to these, the relabeling of generation indices 
of $\nu_{R_i}$ in both $M_\nu$ and $M_R$ gives mass textures which
have the same physical consequences. It is easily shown that all the
above combinations lead to the Majorana mass matrix $M_L$ of the form
(\ref{minML}). Thus from the discussion in Section~\ref{sec:minML},
the solar angle $\theta_{12}$ is found to be larger than the observed
value, when the neutrino mass matrices are combined with the minimal
mass matrices of charged fermions (\ref{uniqueMc}) in unified theory.

Consequently, we are lead to the minimal mass matrices of neutrinos in
the seesaw scenario: $M_\nu$ and $M_R$, which are compatible with
the charged-fermion sector (\ref{uniqueMc}), contain 9 vanishing
elements. There are 3 classes of such neutrino matrices:
\begin{itemize}
\item $M_\nu$ with 5 zeros + $M_R$ with symmetric 4 zeros
\item $M_\nu$ with 6 zeros + $M_R$ with symmetric 3 zeros
\item $M_\nu$ with 7 zeros + $M_R$ with symmetric 2 zeros
\end{itemize}
We exhausted all possibilities of these classes and found a variety of
successful neutrino mass textures. In the case of 5-zero $M_\nu$ and
4-zero $M_R$, 12 patterns of mass textures are consistent with all the
observables for leptons as well as 
quarks (\ref{massexp})$-$(\ref{sin2b}) and (\ref{Nexp}) in
collaboration with the minimal mass matrices (\ref{uniqueMc}). Note
that the result includes 2 patterns of $M_\nu$ and $M_R$ which do not
lead to the minimal form of $M_L$ (\ref{nxtminML}) but successfully
describe all the observables. In the case of 6-zero $M_\nu$ and 
3-zero $M_R$, we found 25 available mass textures and, for 
7-zero $M_\nu$ and 2-zero $M_R$, only 6 patterns are viable. It may be
interesting to note that each texture predicts a definite pattern of
neutrino mass spectrum: there is no matrix form which explains the
observed data both with the normal and inverted mass hierarchies.

The list of the minimal neutrino mass matrices in the seesaw scenario
are summarized in the appendix. We also present for each solution the
resultant Majorana mass matrix $M_L$, possible mass spectrum, and some
phenomenological implications, which might play a key role for the
matrices to be testified.

\section{Phenomenological implications of minimal mass matrices}
In the previous section, we have found considerable patterns of
seesaw-type mass textures compatible with the current experimental
data. To reduce such a wide possibility is an issue of great
importance to our purpose to find the minimal form of mass matrices,
which might reveal a deeper understanding of the structure of Yukawa
couplings. A possible physical difference among the matrix patterns
would appear in yet unknown phenomena involving right-handed
neutrinos. In the following, we discuss flavor-violating decay of
charged leptons and the baryon asymmetry of the universe via thermal
leptogenesis for typical examples of mass matrices in order to clarify
some physical differences among the viable seesaw mass textures.

\subsection{Flavor-violating decay of charged leptons}
Flavor-changing couplings in the lepton Yukawa sector generate the
observable sign of neutrino flavor oscillations. Other physical
effects are possible in physics beyond the standard model and helpful
to discriminate the patterns of Yukawa matrices. A well-known example
we will discuss below is that, if the theory is supersymmetrized,
flavor violation in Yukawa couplings is translated to off-diagonal
components of scalar masses through radiative corrections. That could
induce sizable rates of flavor-changing rare decays for charged
leptons~\cite{LFV} even with the flavor-universal condition for
supersymmetry-breaking couplings motivated by the minimal supergravity
theory.

There are three types of supersymmetry-breaking parameters relevant to
the analysis: the mass parameters of gauge 
fermions $M_{\lambda_a}$ ($a=1,2,3$), the mass and trilinear couplings
of scalar partners of quarks and 
leptons, $(m^2_x)_{ij}$ and $(a_y)_{ij}$ ($i,j=1,2,3$). The latter two
couplings have generation indices $i,j$ and would be additional
sources of flavor violation. In the following, we take, as a 
conservative assumption, the flavor-universal boundary conditions at
some high-energy scale $\Lambda$: $(m^2_x)_{ij}=m_0^2\,\delta_{ij}$ and
$(a_y)_{ij}=a_0\,\delta_{ij}$. However the flavor universality is
disturbed at quantum level by loop corrections involving scalar quarks
and leptons. For example, the renormalization-group running 
from $\Lambda$ down to the right-handed neutrino mass scale induces
off-diagonal elements of left-handed scalar lepton masses:
\begin{equation}
  \big(m^2_{\tilde l}\big)_{ij} \,\sim\,
  \frac{1}{8\pi^2}(3m_0^2 + a_0^2)
  \sum_k (\overline Y_\nu^\dagger)_{ik}(\overline Y_\nu)_{kj}
  \ln\left(\!\frac{\Lambda}{M_{R_k}}\!\right), \qquad (i\neq j).
\end{equation}
The neutrino Yukawa matrix $\overline Y_\nu$ is evaluated in the
generation basis where the charged lepton Yukawa matrix $Y_e$ and
right-handed Majorana mass matrix $M_R$ are flavor 
diagonal [${M_R}_i$ are the (positive) eigenvalues of $M_R$]. With
non-vanishing off-diagonal elements at hand, flavor-violating
processes such as $\mu\to e\gamma$ and $\mu\to e$ conversion in nuclei
generally occurs through the mediation of scalar leptons in loop 
diagrams. The branching ratio for the $\mu\to e\gamma$ decay is
usually dominated by the gaugino-higgsino mixing diagram, and is
roughly estimated as 
\begin{equation}
  {\rm Br}(\mu\to e\gamma) \,\simeq\, \frac{3\alpha}{2\pi}\cdot
  \frac{m_W^4\big|\big(m_{\tilde l}^2\big)_{12}\big|^2}{M_{\rm susy}^8}
  \tan^2\beta,
\end{equation}
where $\alpha$ is the fine structure constant, 
and $m_W$ and $M_{\rm susy}$ denote the mass scales of the $W$ boson
and typical superparticles in the loops, respectively. The decay rate
is enhanced by $\tan\beta$, the ratio of vacuum expectation values of
two Higgs doublets in minimal supersymmetric models. In the numerical
evaluation below, we will also include all other contributions to 
the $\mu\to e\gamma$ decay amplitude. The flavor-violating decay of
the muon might be therefore observed if sizable flavor-changing
supersymmetry-breaking couplings are generated through radiative
corrections which depend on the structure of lepton Yukawa matrices.

For illustrative examples, we take two types of minimal mass textures,
which have been found in the previous section to be consistent with
the current experimental data, and study their predictions for 
charged lepton rare decay, in particular, for $\mu\to e\gamma$. The
first example is a combination of 
textures (see Table~\ref{tab:Mn5MR4} in the appendix),
\begin{equation}
  Y_\nu \,=\, 
  \begin{pmatrix}
    p & q & \\
    & r & \\
    & & s
  \end{pmatrix},\qquad
  M_R \,=\, 
  \begin{pmatrix}
    & t & \\
    t & & \\
    & & u
  \end{pmatrix}.
  \label{example1}
\end{equation}
All the parameters can be made real and positive by phase rotations of
neutrino fields. After integrating out heavy right-handed neutrinos,
one obtains effective Majorana mass matrix for light neutrinos:
\begin{equation}
  M_L \,=\, 
  -\begin{pmatrix}
    & pr/t & \\
    pr/t & 2qr/t & \\
    & & s^2/u
  \end{pmatrix}{\bar v}^2,
\end{equation}
where $\bar v$ is a vacuum expectation value of the Higgs field which
gives neutrino Dirac masses. This matrix $M_L$ takes one of the
minimal forms found in Section~\ref{sec:nxtminML} [the
matrix (\ref{nxtminML1})]. According to the physical predictions
presented there, we find that several (combinations of) parameters are
fixed by the observables:
\begin{equation}
  \frac{p}{q} \,\simeq\, \tan2\theta_{12}, \qquad
  \frac{2r\bar v^2}{t}\sqrt{pq} \,\simeq\, 
  \sqrt{\Delta m_{21}^2\sin2\theta_{12}}, \qquad
  \frac{(s\bar v)^2}{u} \,\simeq\, \sqrt{\Delta m_{32}^2},
  \label{parafix}
\end{equation}
neglecting $\mathcal{O}(10^{-2})$ corrections. The atmospheric
neutrino oscillation is explained by the charged lepton sector, and
leads to no significant restriction on the neutrino side. With the
textures (\ref{example1}), a pair of off-diagonal elements of scalar
lepton mass matrix is radiatively generated through
renormalization-group running, which is found to be proportional to
\begin{equation}
  (\overline Y_\nu^\dagger\overline Y_\nu^{})_{21} \,\simeq\,
  \frac{1}{\sqrt{2}}\Big[\,pq+(-p^2+q^2+r^2)(V_{eL})_{21}
  +(p^2-s^2)(V_{eL})_{31}\Big].
  \label{YY21-1}
\end{equation}
Notice that, since the first-two generation right-handed neutrinos are
degenerate for $M_R$ (\ref{example1}), the basis rotation which
diagonalizes $M_R$ is canceled out in the 
expression (\ref{YY21-1}). In the limit $p=r=s$, the experimental
relations (\ref{parafix}) are still satisfied with appropriate values
of other parameters, while the off-diagonal element (\ref{YY21-1})
does not vanish and gives rise to a unsuppressed decay rate 
of $\mu\to e\gamma$ proportional to $q^2$.

\begin{figure}[t]
\centering
\includegraphics[scale=1.5]{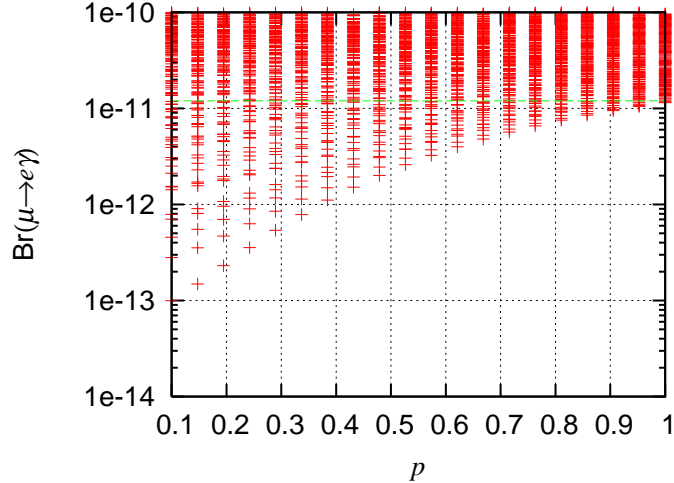}
\caption{A plot of the branching ratio of $\mu\to e\gamma$ for the
texture example (\ref{example1}). The horizontal axis denotes a Yukawa
coupling constant. The horizontal (green) line shows the current
experimental upper bound of the branching ratio. The universal scalar
mass parameter and gaugino masses are taken to 
be $m_0=M_{\lambda_1}=M_{\lambda_2}=M_{\lambda_3}=800$ GeV 
and $a_0=0$ at the unification scale. The branching ratio is
scaled with $\tan^2\beta$, and here we assume $\tan\beta=10$.\bigskip}
\label{fig:meg1}
\end{figure}

Figure~\ref{fig:meg1} shows a numerical result of 
the $\mu\to e\gamma$ decay rate for the texture 
example (\ref{example1}). In this figure, the neutrino Yukawa
couplings $p$, $r$, and $s$ vary between $0.1$ and $1$, while the
other parameters $q$, $t$, and $u$ are fixed so that the neutrino
oscillation data is reproduced. We assume, as an example, the
universal mass parameter $m_0$ and gaugino 
masses $M_{\lambda_{1,2,3}}$ to be 800 GeV at the unification scale,
and $\tan\beta=10$. It can be seen from the figure that, for larger
values of Yukawa couplings, the flavor structure (\ref{example1}) is
already excluded by the absence of experimental signature of
charged lepton rare decay.

Another texture example we consider is a combination (see
Table~\ref{tab:Mn6MR3} in the appendix),
\begin{equation}
  Y_\nu \,=\, 
  \begin{pmatrix}
    & p & \\
    q & & \\
    & & r
  \end{pmatrix},\qquad
  M_R \,=\, 
  \begin{pmatrix}
    & s & \\
    s & t & \\
    & & u
  \end{pmatrix}.
  \label{example2}
\end{equation}
All the matrix elements can be made real and positive. It may be
interesting to note that, among the minimal forms of neutrino mass
matrices presented in the appendix, only this texture example leads to
the same form of $M_L$ as the previous one (\ref{example1}). This is
because the minimal mass matrices (\ref{example2}) are chosen for the
second example as a comparison. In the present case, the induced
Majorana mass matrix for light neutrinos is
\begin{equation}
  M_L \,=\, 
  -\begin{pmatrix}
    & pq/s & \\
    pq/s & -p^2t/s^2 & \\
    & & r^2/u
  \end{pmatrix}{\bar v}^2,
\end{equation}
which is of the identical form to the first example and gives the same
predictions for neutrino masses and mixing angles. However the flavor
structure of neutrino Yukawa matrix is different from (\ref{YY21-1}):
\begin{equation}
  (\overline Y_\nu^\dagger\overline Y_\nu^{})_{21} \,\simeq\, 
  \frac{1}{\sqrt{2}}\Big[\,(p^2-q^2)(V_{eL})_{21} 
  +(q^2-r^2)(V_{eL})_{31}\Big],
  \label{YY21-2}
\end{equation}
which contains 3 free parameters $p$, $q$, and $r$. The other matrix
elements are determined by the physical observables similarly 
to (\ref{parafix}). The off-diagonal element (\ref{YY21-2}) describes
induced flavor-violating masses of scalar leptons approximately in the
situation that there is no large mass hierarchy among the right-handed
neutrinos. Note, in the present case, that one can take the 
limit $p=q=r$ without being incompatible with the experimental data,
which limit suppresses the amplitude of $\mu\to e\gamma$ up to tiny
contribution from right-handed scalar lepton masses. Thus the textures
(\ref{example2}) generally tend to predict smaller branching ratio of
charged lepton rare decay even with large neutrino Yukawa couplings
of $\mathcal{O}(1)$.

\begin{figure}[t]
\centering
\includegraphics[scale=1.5]{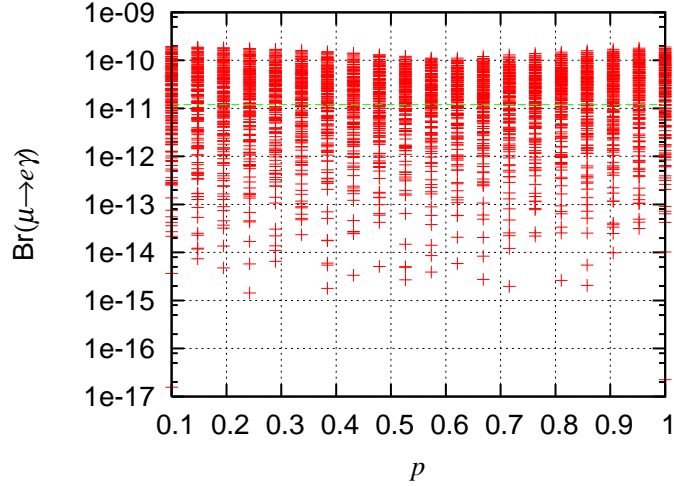}
\caption{A plot of the branching ratio of $\mu\to e\gamma$ for the
texture example (\ref{example2}). The horizontal axis denotes a Yukawa
coupling constant. The horizontal (green) line shows the current
experimental upper bound of the branching ratio. The universal scalar
mass parameters and gaugino masses are taken at the unification scale
to be the same as in Figure~\ref{fig:meg1}. Here we 
assume $\tan\beta = 10$.\bigskip}
\label{fig:meg2}
\end{figure}

In Figure~\ref{fig:meg2}, we present a numerical result of 
the $\mu\to e\gamma$ decay rate for the texture 
example (\ref{example2}). In this figure, the neutrino Yukawa
couplings $p$, $q$, and $r$ vary between $0.1$ and $1$. The
supersymmetry-breaking parameters are taken to be the same as in
Figure~\ref{fig:meg1}. The figure shows that the branching ratio can
be suppressed ($<10^{-12}$) even for a relatively large Yukawa
coupling ($0.4<p<1$), which fact makes a sharp contrast with the first
texture example (see Figure~\ref{fig:meg1}).

\smallskip

In this way, the measurement of flavor-changing rare decay of charged
leptons would be able to split the degeneracy of possible texture
forms of neutrinos.

\subsection{Thermal leptogenesis}
Another interesting physical effect of neutrino mass textures comes
from the existence of high-energy CP violation in the Yukawa
couplings. That is, in some classes of texture forms, there remain
complex phases of couplings which cannot be rotated away by the
redefinition of neutrino fields. Such phase degrees of freedom may
explain the baryon asymmetry of the universe through the thermal 
leptogenesis~\cite{leptogen}.

Let us consider the following texture form, which is one of the
minimal neutrino mass matrices found in Section~\ref{sec:minMnMR}
(see Table~\ref{tab:Mn5MR4} in the appendix):
\begin{equation}
  Y_\nu \,=\,
  \begin{pmatrix}
    pe^{i\phi} & & \\
    q & \!\!r & \\
    & \!\!s & ~~~
  \end{pmatrix}, \qquad
  M_R \,=\, 
  \begin{pmatrix}
    & t & \\
    t & & \\
    & & u\,
  \end{pmatrix}.
  \label{example3}
\end{equation}
To study CP-violating effects, it is convenient to reduce the phase
degrees of freedom in the Yukawa couplings. In the above matrix forms,
we have taken a basis where all the parameters but $(Y_\nu)_{11}$ are
real (and positive) by redefining neutrino fields. The resultant
Majorana mass matrix for light neutrinos becomes
\begin{equation}
  M_L \,=\,
  -\begin{pmatrix}
    2pqe^{2i\phi}/t & pr/t & \\
    pr/t & s^2/u & \\
    & & ~~~~
  \end{pmatrix}{\bar v}^2.
\end{equation}
This form of $M_L$ predicts the inverted mass hierarchy of neutrinos,
detectable neutrino-less double beta decay, and low-energy
CP-violation in future neutrino oscillation experiments which is found
to be proportional 
to $\sin(\phi+\phi_e)\cos\phi$, where $\phi_e$ originates from
unremovable phases of charged lepton Yukawa couplings.

The CP asymmetry generated in the decay of the $i$-th generation
right-handed neutrino is described by the quantity
\begin{equation}
  \epsilon_i \;=\; \frac{\sum_j\Gamma(\nu_{R_i}\to L_jH)
    -\sum_j\Gamma(\nu_{R_i}\to L^c_jH^\dagger)}{\sum_j
    \Gamma(\nu_{R_i}\to L_jH) 
    +\sum_j\Gamma(\nu_{R_i}\to L^c_jH^\dagger)},
\end{equation}
and the resultant leptonic asymmetry $\eta_L$~\cite{KT} is evaluated as
\begin{equation}
  \eta_L \,\simeq\, \frac{1}{g_*}\sum_i\kappa_i\epsilon_i,
\end{equation}
where $\kappa_i$ are the factors which represent wash-out effects
governed by the Boltzmann equations, and $g_*$ is the number of
relativistic degrees of freedom. 
An approximate analytic expression for the factor $\kappa_i$ is given by
\begin{equation}
\kappa_i \quad\simeq\quad 0.3 \left( \frac{10^{-3} \mathrm{eV}}{
    \tilde{m}_i }\right)\left(\ln \frac{\tilde{m}_i}{10^{-3}
    \mathrm{eV} }\right)^{-0.6},
\quad\quad \tilde{m}_i \equiv (M_\nu
M_\nu^\dag)_{ii}/M_{R_i},
\label{dilution}
\end{equation}
if right-handed neutrinos are enough strongly coupled to
the thermal bath: $ \tilde{m}_i > 10^{-3}$ eV.
The electroweak anomaly which is in
thermal equilibrium at the temperature $10^2\sim10^{12}$ GeV converts
a part of $B-L$ asymmetry into $B$ asymmetry in such a way 
that $\eta_B=-28/51\times\eta_L$ in the standard model~\cite{LtoB}.

In addition to the existence of CP-violating phase, the matrix forms
(\ref{example3}) are also interesting in that the right-handed
neutrino masses are degenerate in the first and second
generations. Such a degeneracy is known to enhance the leptonic
asymmetry with self-energy contribution in the decay of right-handed
neutrinos~\cite{resonant}. That helps to achieve the observed range of
the baryon asymmetry with relatively light right-handed neutrinos,
which is reasonable from a viewpoint of the cosmological problem of
unstable gravitino in supersymmetric theory~\cite{gravitino}. We
discuss below the possibility of such enhancement of lepton asymmetry
for the textures (\ref{example3}).

In the region where $\nu_{R_1}$ and $\nu_{R_2}$ are nearly degenerate
in mass, the CP-asymmetry parameters are well approximated
as~\cite{resonant}
\begin{equation}
  \epsilon_1 \,\simeq\, \frac{{\rm Im}\,A^2_{12}}{A_{11}}\cdot
  \frac{R}{R^2+A^2_{22}}, \qquad
  \epsilon_2 \,\simeq\, \frac{{\rm Im}\,A^2_{12}}{A_{22}}\cdot
  \frac{R}{R^2+A^2_{11}},
  \label{CPasym}
\end{equation}
where $A_{ij}=(Y_\nu Y_\nu^\dagger)_{ij}/8\pi$ and 
$R=(M_{R_1}^2-M_{R_2}^2)/M_{R_1}M_{R_2}$. The mass eigenvalues of
right-handed neutrinos $M_{R_i}$ are taken to be real and
positive. It is noticed that there exist the regulating 
terms, $A_{11}^2$ and $A_{22}^2$, which come from the resummation of
self-energy diagrams.

\begin{figure}[t]
\centering
\includegraphics[scale=1.2]{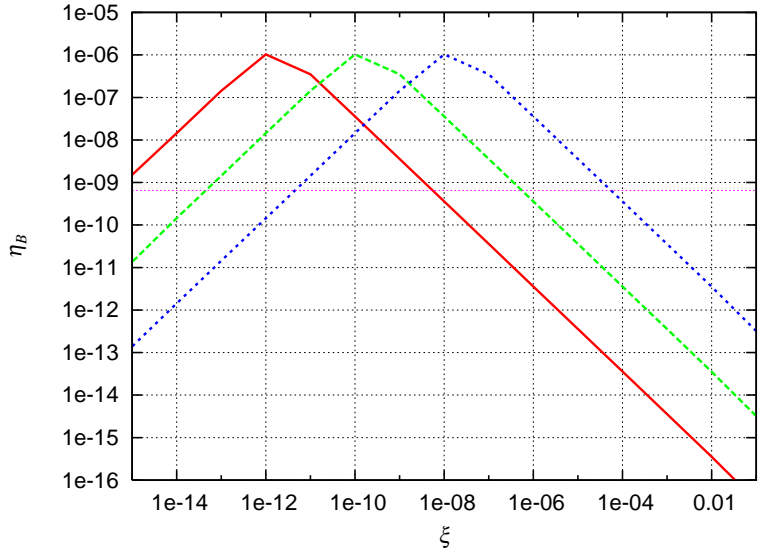}
\caption{The baryon-to-photon ratio as the function of the degeneracy
parameter $\xi$ for one of the minimal forms of seesaw mass matrices  
[(\ref{example3})]. The horizontal dotted line shows the observed
value of baryon asymmetry. The red-solid, green-dashed, and
blue-dotted lines show the predictions 
for $(p,M_{R_1})=$ ($10^{-5},10^5$ GeV), ($10^{-4},10^7$ GeV), 
and ($10^{-3},10^9$ GeV), respectively. For other parameters, we take 
$M_{R_2}=(1-\xi)M_{R_1}$, $M_{R_3}=10^2M_{R_1}$,
$m_1=4.61\times10^{-2}$ eV, $m_2=4.69\times10^{-2}$ eV,
$\theta_{12}=33^\circ$, and $\phi=\pi/2$.\bigskip}
\label{fig:etaB}
\end{figure}

If the degeneracy of two heavy Majorana neutrinos is exact ($R\to0$),
the asymmetry parameters (\ref{CPasym}) vanish. In the present case we
are discussing, the matrix $M_R$ (\ref{example3}) means the equal mass
eigenvalues for the first-two right-handed neutrinos. However such an
exact degeneracy may be relaxed, for example, by radiative corrections
from other sectors such as $Y_\nu$. In fact, a required order of mass
splitting is generally small and could easily be realized. We define
the parameter $\xi$ to describe the degree of degeneracy 
for $M_{R_1}$ and $M_{R_2}$:
\begin{equation}
  \frac{M_{R_2}}{M_{R_1}} \,=\, 1-\xi.
\end{equation}
Using (\ref{dilution}) and (\ref{CPasym}), we can calculate the 
baryon asymmetry.
The predicted baryon asymmetry as the function of $\xi$ is shown in
Figure~\ref{fig:etaB}.
The red-solid, green-dashed, and blue-dotted lines show the
predictions for $M_{R_1}\simeq M_{R_2}=10^5$, $10^7$, and $10^9$~GeV,
respectively. The other masses and Yukawa coupling constants are
determined so that the results from neutrino oscillation experiments
are satisfied. As seen from the figure, the observed baryon asymmetry
of the universe $\eta_B=(6.2-6.9)\times10^{-10}$~\cite{WMAP} is
reproduced for each parameter set with a small mass 
splitting $\xi$. Recent studies on the decay of gravitino suggest that
the reheating temperature of the universe is preferred to be less 
than $10^7$~GeV for successful big-bang nucleosynthesis with the 
unstable gravitino lighter than $10$~TeV~\cite{gravitino}. We here
found that the structure of mass 
matrices (\ref{example3}) is capable of explaining the observed baryon
asymmetry while solving the cosmological gravitino problem. It would
be an interesting task to analyze the other minimal forms of neutrino
mass matrices presented in the appendix whether they can reproduce the
proper value of $\eta_B$ in the leptogenesis scenario.

\smallskip

Finally we comment on the leptonic CP violation. In
Figure~\ref{fig:etaB}, we have set the unremovable complex phase in 
the Yukawa couplings as $\phi=\pi/2$. This does not lead to low-energy
CP violation in future neutrino oscillation experiments, which is
proportional to $\cos\phi$. However that is just an illustrative
example, and the detailed analysis of CP-violating effects
discriminating possible forms of minimal lepton mass matrices is left
to future investigation.

\section{Summary}
In this paper, we have studied mass matrices of quarks and leptons in
unified theory which have as many vanishing elements as
possible. First we have found that the following form of mass matrices
is minimal and consistent with the current experimental data of quarks
and charged leptons, including the recent measurement of CP violation
in the B-meson system:
\begin{equation}
  M_u \,=\, \begin{pmatrix}
    & a & \\
    a & & b \\
    & b & c
  \end{pmatrix},\quad\;
  M_d \,=\, \begin{pmatrix}
    & e & \\
    d & & h \\
    & g & f
  \end{pmatrix},\quad\;
  M_e \,=\, \begin{pmatrix}
    & e & \\
    d & & \!\!-3h \\
    & g & f
  \end{pmatrix}.
  \label{ude}
\end{equation}
Trivial exchanges of generation labels of $Q_i$ and $u_{R_i}$ do not
change physical predictions and lead to viable solutions. An
interesting observation in the matrix form (\ref{ude}) is that it
induces a large mixture of left-handed charged leptons between the
second and third generations, which is suitable to the experimental
results for the atmospheric neutrinos.

Bases on this matrix form of charged fermions, we have investigated
two types of mass matrices for neutrinos. The first analysis deals
with the effective Majorana mass matrix $M_L$ of left-handed
neutrinos. We have found that the following forms are 
minimal, i.e.\ contain the maximal number of vanishing elements:
\begin{equation}
  M_L \;=\;
  \begin{pmatrix}
    & l & \\
    l & m & \\
    & & n
  \end{pmatrix}, \;
  \begin{pmatrix}
    & l & \\
    l & & m \\
    & m & n
  \end{pmatrix}, \;
  \begin{pmatrix}
    l & m & \\
    m & n & \\
    & & ~
  \end{pmatrix}, \;
  \begin{pmatrix}
    & & l \\
    & & m \\
    l & m & n
  \end{pmatrix}.
  \label{l}
\end{equation}
Combined with (\ref{ude}), all current experimental results of quarks
and leptons are explained. Trivial exchanges of generation labels 
of $L_i$ (and $d_{R_i}$ in unified theory) in (\ref{ude}) and
(\ref{l}) do not change physical predictions and lead to viable 
solutions. In addition, the exchange of the second and third labels 
in $M_L$ also gives the matrices experimentally allowed, because of a
large mixture of left-handed charged leptons between the second and
third generations mentioned above. Each set of minimal mass textures
predicts typical mass spectrum and generation mixing of neutrinos
(Table~\ref{tab:predict}), and is therefore testable with precise
measurements of masses and mixing angles in future neutrino experiments.

The second type of our analysis introduces heavy right-handed
neutrinos, and discuss the minimality of neutrino Dirac mass 
matrix $M_\nu$ and Majorana mass matrix $M_R$ of right-handed
neutrinos. In this case, we have found a variety of successful mass
textures presented in the appendix, which are classified into three
categories:
\begin{itemize}
\item $M_\nu$ with 5 zeros + $M_R$ with symmetric 4 zeros
~(12 patterns)
\item $M_\nu$ with 6 zeros + $M_R$ with symmetric 3 zeros
~(25 patterns)
\item $M_\nu$ with 7 zeros + $M_R$ with symmetric 2 zeros
~(6 patterns)
\end{itemize}
Integrated out heavy right-handed neutrinos, almost all the patterns
effectively induce $M_L$ in the forms of (\ref{l}), but the other 5
patterns generate different forms of $M_L$.

As typical phenomenological implications of the minimal mass matrices
of neutrinos, we have discussed the $\mu\to e\gamma$ decay and the
thermal leptogenesis scenario. In the study of $\mu\to e\gamma$, some
of textures are found to predict large decay amplitudes in natural
parameter regions to reach the experimental upper bound. It also turns
out that mass matrices with unremovable complex phases of matrix
elements can explain the observed baryon asymmetry of the universe via
thermal leptogenesis. In particular, some types of textures avoid an
apparent inconsistency between viable thermal leptogenesis and the
gravitino problem in supersymmetric theory by virtue of resonant
behavior in the decays of right-handed neutrinos. Future neutrino
experiments for low-energy leptonic CP violation and neutrino-less
double beta decay would be expected to distinguish the minimal forms
of quark and lepton mass matrices.

\bigskip\bigskip
\subsection*{Acknowledgments}
 
The authors thank Nobuhiro Uekusa for collaboration at an early stage
of this work. A.W.\ is grateful to the organizers and participants of
the workshop ``Summer Institute 2005'' held at Fuji-Yoshida,
Yamanashi, Japan on 11-18 August 2005, for useful discussions to
complete the work. This work is partially supported by scientific
grant from the Ministry of Education, Science, Sports, and Culture of
Japan (No.~17740150) and by grant-in-aid for the scientific research
on priority area "Progress in elementary particle physics of the 21st
century through discoveries of Higgs boson and supersymmetry"
(No.~16081209).

\newpage
\section*{Appendix. ~The minimal forms of seesaw mass textures}
In this appendix, we present the minimal forms of neutrino mass
matrices in the seesaw scenario (see Section~\ref{sec:minMnMR}). These
matrices are consistent with the current experimental data of quarks
and leptons (\ref{massexp})$-$(\ref{sin2b}) and (\ref{Nexp}) in
collaboration with the minimal solution (\ref{uniqueMc}) for charged
fermions.

The following tables show experimentally viable forms 
of $M_\nu$ and $M_R$. It should be noted that, in addition to the
solutions presented in the tables, there are two types of additional
matrix forms which are obtained by exchanging generation indices (rows
and/or columns). The first is the label exchange of right-handed
neutrinos. In the seesaw formula, the simultaneous exchange of the
same generation indices of $\nu_{R_i}$ in $M_\nu$ and $M_R$ does not
affect physical consequences at low-energy regime. Secondly, since the
charged lepton mixing matrix $V_{eL}$ is approximately invariant under
the exchange of the second and third 
rows ($\nu_{L_2}\leftrightarrow \nu_{L_3}$), the same exchange 
in $M_\nu$ also gives available matrix forms.

In the tables, we show several results for each combination of
neutrino mass matrices: ``$M_L$'', ``spectrum'', ``CP'', 
and ``$0\nu\beta\beta$''. The column $M_L$ shows the mass matrix forms
of left-handed neutrinos effectively induced after the seesaw
operation. In the column ``spectrum'', we present possible mass
patterns of light neutrinos for each texture. The abbreviation NH
denotes the normal mass hierarchy $m_1< m_2\ll m_3$ and IH the
inverted one $m_3\ll m_1<m_2$. The column ``CP'' shows that matrices
contain unremovable complex phases, which may be required to implement
the leptogenesis and also leads to low-energy CP violation in neutrino
oscillation experiments. The symbol $\bigcirc$ means that there exist
phase degrees of freedom in $M_\nu$ and $M_R$ which cannot be rotated
away by the redefinition of neutrino fields, and $\times$ means the
absence of such complex phases. Finally, for ``$0\nu\beta\beta$'', the
symbol $\bigcirc$ ($\times$) implies that the averaged 
mass $|\langle m_{ee}\rangle|$ is larger (smaller) 
than $10^{-2}$ eV\@. The averaged neutrino mass 
for $0\nu\beta\beta$ is defined for the parametrization of the MNS
matrix given in Section~\ref{sec:exp} as
$\langle m_{ee}\rangle=
m_1\cos^2\theta_{12}\cos^2\theta_{13}e^{-2i\rho}+
m_2\sin^2\theta_{12}\cos^2\theta_{13}e^{-2i\sigma}+
m_3\sin^2\theta_{13}e^{2i\delta}$.

\newpage

\begin{table}[htbp]
\renewcommand{\arraystretch}{1.05}
\centering
\begin{tabular}{cccccccc} \hline\hline
$M_\nu$ & $M_R$ &&& $M_L$ & spectrum & CP & $0\nu\beta\beta$ \\ \hline
$\begin{pmatrix}
  p & q & \\
  & r & \\
  & & s
\end{pmatrix}$ &
$\begin{pmatrix}
  & t & \\
  t & & \\
  & & u
\end{pmatrix}$
&&$\to$&
$\begin{pmatrix}
  & l & \\
  l & m & \\
  & & n
\end{pmatrix}$
& NH & $\times$ & $\times$ \\ \hline
$\begin{pmatrix}
  & & p \\
  & q & \\
  r & & s
\end{pmatrix}$ &
$\begin{pmatrix}
  t & & \\
  & & u \\
  & u &
\end{pmatrix}$ 
&&$\to$&
$\begin{pmatrix}
  & l & \\
  l & & m \\
  & m & n
\end{pmatrix}$
& NH & $\times$ & $\times$ \\ \hline
$\begin{pmatrix}
  p & & \\
  q & r& \\
  & s & ~
\end{pmatrix}$ &
$\begin{pmatrix}
  & t & \\
  t & & \\
  & & u
\end{pmatrix}$
&&$\to$&
$\begin{pmatrix}
  l & m & \\
  m & n & \\
  & & ~
\end{pmatrix}$
& IH & $\bigcirc$ & $\bigcirc$ \\ \hline
$\begin{pmatrix}
  p & & ~ \\
  q & r & \\
  & s &
\end{pmatrix}$ &
$\begin{pmatrix}
  t & & \\
  & & u \\
  & u &
\end{pmatrix}$
&&$\to$&
$\begin{pmatrix}
  l & m & \\
  m & n & \\
  & & ~
\end{pmatrix}$
& IH & $\bigcirc$ & $\bigcirc$ \\ \hline
$\begin{pmatrix}
  p & & ~ \\
  q & r & \\
  & s &
\end{pmatrix}$ &
$\begin{pmatrix}
  & & t \\
  & u & \\
  t & & 
\end{pmatrix}$
&&$\to$&
$\begin{pmatrix}
  l & m & \\
  m & n & \\
  & & ~
\end{pmatrix}$
& IH & $\bigcirc$ & $\bigcirc$ \\ \hline
$\begin{pmatrix}
  p & q & \\
  & r & \\
  & s & ~
\end{pmatrix}$ &
$\begin{pmatrix}
  t & & \\
  & & u \\
  & u &
\end{pmatrix}$
&&$\to$&
$\begin{pmatrix}
  l & m & \\
  m & n & \\
  & & ~
\end{pmatrix}$
& IH & $\bigcirc$ & $\bigcirc$ \\ \hline
$\begin{pmatrix}
  p & q & \\
  r & & \\
  s & & ~
\end{pmatrix}$ &
$\begin{pmatrix}
  t & & \\
  & & u \\
  & u &
\end{pmatrix}$
&&$\to$&
$\begin{pmatrix}
  l & m & \\
  m & n & \\
  & & ~
\end{pmatrix}$
& IH & $\bigcirc$ & $\bigcirc$ \\ \hline
$\begin{pmatrix}
  & & \\
  & & p \\
  q & r & s
\end{pmatrix}$ &
$\begin{pmatrix}
  t & & \\
  & & u \\
  & u &
\end{pmatrix}$
&&$\to$&
$\begin{pmatrix}
  & & l \\
  & & m \\
  l & m & n
\end{pmatrix}$
& NH & $\times$ & $\times$ \\ \hline
$\begin{pmatrix}
  & & \\
  p & & q \\
  & r & s
\end{pmatrix}$ &
$\begin{pmatrix}
  t & & \\
  & & u \\
  & u &
\end{pmatrix}$
&&$\to$&
$\begin{pmatrix}
  & l & m \\
  l & & n \\
  m & n & \!\!\frac{2mn}{l}
\end{pmatrix}\!$
& NH & $\times$ & $\times$ \\ \hline
$\begin{pmatrix}
  p & q & \\
  r & s & \\
  & &
\end{pmatrix}$ &
$\begin{pmatrix}
  t & & \\
  & & u \\
  & u &
\end{pmatrix}$
&&$\to$&
$\begin{pmatrix}
  l & m & \\
  m & n & \\
  & & ~
\end{pmatrix}$
& IH & $\times$ & $\bigcirc$ \\ \hline
$\begin{pmatrix}
  p & q & \\
  & & r \\
  & & s
\end{pmatrix}$ &
$\begin{pmatrix}
  t & & \\
  & & u \\
  & u &
\end{pmatrix}$
&&$\to$&
$\begin{pmatrix}
  \frac{l^2}{m} & l & \\
  l & m & \\
  & & n
\end{pmatrix}$
& NH & $\times$ & $\times$ \\ \hline
$\begin{pmatrix}
  p & q & \\
  & & r \\
  & & s
\end{pmatrix}$ &
$\begin{pmatrix}
  & & t \\
  & u & \\
  t & &
\end{pmatrix}$
&&$\to$&
$\begin{pmatrix}
  & & l \\
  & & m \\
  l & m & n
\end{pmatrix}$
& NH & $\times$ & $\times$ \\ \hline\hline
\end{tabular}
\caption{$M_\nu$ with 5 zeros + $M_R$ with symmetric 4 zeros.}
\label{tab:Mn5MR4}
\end{table}

\newpage

\begin{table}[htbp]
\renewcommand{\arraystretch}{1.05}
\centering
\begin{tabular}{cccccccc} \hline\hline
$M_\nu$ & $M_R$ &&& $M_L$ & spectrum & CP & $0\nu\beta\beta$ \\ \hline
$\begin{pmatrix}
  & p & \\
  q & & \\
  & & r
\end{pmatrix}$ &
$\begin{pmatrix}
  & s & \\
  s & t & \\
  & & u
\end{pmatrix}$
&&$\to$&
$\begin{pmatrix}
  & l & \\
  l & m & \\
  & & n
\end{pmatrix}$ 
& NH & $\times$ & $\times$ \\ \hline
$\begin{pmatrix}
  & p & ~ \\
  q & & \\
  r & &
\end{pmatrix}$ &
$\begin{pmatrix}
  & s & \\
  s & t & \\
  & & u
\end{pmatrix}$
&&$\to$&
$\begin{pmatrix}
  l & m & \\
  m & n & \\
  & & ~
\end{pmatrix}$
& IH & $\times$ & $\bigcirc$ \\ \hline
$\begin{pmatrix}
  p & & ~ \\
  & q & \\
  & r &
\end{pmatrix}$ &
$\begin{pmatrix}
  & s & \\
  s & t & \\
  & & u
\end{pmatrix}$
&&$\to$&
$\begin{pmatrix}
  l & m & \\
  m & n & \\
  & & ~
\end{pmatrix}$
& IH & $\times$ & $\bigcirc$ \\ \hline
$\begin{pmatrix}
  p & & \\
  q & & \\
  & r & ~
\end{pmatrix}$ &
$\begin{pmatrix}
  & s & \\
  s & & t \\
  & t & u
\end{pmatrix}$
&&$\to$&
$\begin{pmatrix}
  l & m & \\
  m & n & \\
  & & ~
\end{pmatrix}$
& IH & $\times$ & $\bigcirc$ \\ \hline
$\begin{pmatrix}
  & p & \\
  & q & \\
  r & & ~
\end{pmatrix}$ &
$\begin{pmatrix}
  & s & \\
  s & & t \\
  & t & u
\end{pmatrix}$
&&$\to$&
$\begin{pmatrix}
  l & m & \\
  m & n & \\
  & & ~
\end{pmatrix}$
& IH & $\times$ & $\bigcirc$ \\ \hline
$\begin{pmatrix}
  & p & \\
  q & & \\
  r & & ~
\end{pmatrix}$ &
$\begin{pmatrix}
  & s & \\
  s & & t \\
  & t & u
\end{pmatrix}$
&&$\to$&
$\begin{pmatrix}
  l & m & \\
  m & n & \\
  & & ~
\end{pmatrix}$
& IH & $\times$ & $\bigcirc$ \\ \hline
$\begin{pmatrix}
  p & & \\
  & q & \\
  & r & ~
\end{pmatrix}$ &
$\begin{pmatrix}
  & s & \\
  s & & t \\
  & t & u
\end{pmatrix}$
&&$\to$&
$\begin{pmatrix}
  l & m & \\
  m & n & \\
  & & ~
\end{pmatrix}$
& IH & $\times$ & $\bigcirc$ \\ \hline
$\begin{pmatrix}
  & & \\
  p & q & \\
  & & r
\end{pmatrix}$ &
$\begin{pmatrix}
  s & & \\
  & t &  \\
  & & u
\end{pmatrix}$
&&$\to$&
$\begin{pmatrix}
  \frac{l^2}{m} & l & \\
  l & m & \\
  & & n
\end{pmatrix}$
& NH & $\times$ & $\times$ \\ \hline
$\begin{pmatrix}
  & & \\
  p & q & \\
  & & r
\end{pmatrix}$ &
$\begin{pmatrix}
  s & & \\
  & t & u \\
  & u &
\end{pmatrix}$
&&$\to$&
$\begin{pmatrix}
  & & l \\
  & & m \\
  l & m & n
\end{pmatrix}$
& NH & $\times$ & $\times$ \\ \hline
$\begin{pmatrix}
  & & \\
  p & q & \\
  & & r
\end{pmatrix}$ &
$\begin{pmatrix}
  s & t & \\
  t & & u \\
  & u &
\end{pmatrix}$
&&$\to$&
$\begin{pmatrix}
  & & l \\
  & & m \\
  l & m & n
\end{pmatrix}$
& NH & $\times$ & $\times$ \\ \hline
$\begin{pmatrix}
  p & q & \\
  & & \\
  & & r
\end{pmatrix}$ &
$\begin{pmatrix}
  & s & \\
  s & t & \\
  & & u
\end{pmatrix}$
&&$\to$&
$\begin{pmatrix}
  \frac{l^2}{m} & l & \\
  l & m & \\
  & & n
\end{pmatrix}$
& NH & $\times$ & $\times$ \\ \hline
$\begin{pmatrix}
  p & q & \\
  & & \\
  & & r
\end{pmatrix}$ &
$\begin{pmatrix}
  s & & \\
  & t & u \\
  & u &
\end{pmatrix}$
&&$\to$&
$\begin{pmatrix}
  \frac{l^2}{m} & l & \\
  l & m & \\
  & & n
\end{pmatrix}$
& NH & $\times$ & $\times$ \\ \hline
\end{tabular}
\caption{$M_\nu$ with 6 zeros + $M_R$ with symmetric 3 zeros.}
\label{tab:Mn6MR3}
\end{table}

\newpage

\addtocounter{table}{-1}
\begin{table}[htbp]
\renewcommand{\arraystretch}{1.05}
\centering
\begin{tabular}{cccccccc} \hline
$M_\nu$ & $M_R$ &&& $M_L$ & spectrum & CP & $0\nu\beta\beta$ \\ \hline
$\begin{pmatrix}
  p & q & \\
  r & & \\
  & & ~
\end{pmatrix}$ &
$\begin{pmatrix}
  s & & \\
  & t & \\
  & & u
\end{pmatrix}$
&&$\to$&
$\begin{pmatrix}
  l & m & \\
  m & n & \\
  & & ~
\end{pmatrix}$
& IH & $\bigcirc$ & $\bigcirc$ \\ \hline
$\begin{pmatrix}
  p & q & \\
  r & & \\
  & & ~
\end{pmatrix}$ &
$\begin{pmatrix}
  & s & \\
  s & & t \\
  & t & u
\end{pmatrix}$
&&$\to$&
$\begin{pmatrix}
  l & m & \\
  m & n & \\
  & & ~
\end{pmatrix}$
& IH & $\bigcirc$ & $\bigcirc$ \\ \hline
$\begin{pmatrix}
  p & q & \\
  & r & \\
  & & ~
\end{pmatrix}$ &
$\begin{pmatrix}
  & s & \\
  s & & t \\
  & t & u
\end{pmatrix}$
&&$\to$&
$\begin{pmatrix}
  l & m & \\
  m & n & \\
  & & ~
\end{pmatrix}$
& IH & $\bigcirc$ & $\bigcirc$ \\ \hline
$\begin{pmatrix}
  p & & ~ \\
  & & \\
  q & r &
\end{pmatrix}$ &
$\begin{pmatrix}
  & s & \\
  s & & t \\
  & t & u
\end{pmatrix}$
&&$\to$&
$\begin{pmatrix}
  l & m & \\
  m & n & \\
  & & ~
\end{pmatrix}$
& IH & $\bigcirc$ & $\bigcirc$ \\ \hline
$\begin{pmatrix}
  & p & \\
  & & ~ \\
  q & r &
\end{pmatrix}$ &
$\begin{pmatrix}
  & s & \\
  s & & t \\
  & t & u
\end{pmatrix}$
&&$\to$&
$\begin{pmatrix}
  l & m & \\
  m & n & \\
  & & ~
\end{pmatrix}$
& IH & $\bigcirc$ & $\bigcirc$ \\ \hline
$\begin{pmatrix}
  p & q & \\
  & & ~ \\
  r & &
\end{pmatrix}$ &
$\begin{pmatrix}
  & s & \\
  s & & t \\
  & t & u
\end{pmatrix}$
&&$\to$&
$\begin{pmatrix}
  l & m & \\
  m & n & \\
  & & ~
\end{pmatrix}$
& IH & $\bigcirc$ & $\bigcirc$ \\ \hline
$\begin{pmatrix}
  p & q & \\
  & & ~ \\
  & r &
\end{pmatrix}$ &
$\begin{pmatrix}
  & s & \\
  s & & t \\
  & t & u
\end{pmatrix}$
&&$\to$&
$\begin{pmatrix}
  l & m & \\
  m & n & \\
  & & ~
\end{pmatrix}$
& IH & $\bigcirc$ & $\bigcirc$ \\ \hline
$\begin{pmatrix}
  p & q & \\
  r & & \\
  & & ~
\end{pmatrix}$ &
$\begin{pmatrix}
  & s & \\
  s & t & \\
  & & u
\end{pmatrix}$
&&$\to$&
$\begin{pmatrix}
  l & m & \\
  m & n & \\
  & & ~
\end{pmatrix}$
& IH & $\bigcirc$ & $\bigcirc$ \\ \hline
$\begin{pmatrix}
  p & q & \\
  & r & \\
  & & ~
\end{pmatrix}$ &
$\begin{pmatrix}
  & s & \\
  s & t & \\
  & & u
\end{pmatrix}$
&&$\to$&
$\begin{pmatrix}
  l & m & \\
  m & n & \\
  & & ~
\end{pmatrix}$
& IH & $\bigcirc$ & $\bigcirc$ \\ \hline
$\begin{pmatrix}
  p & & \\
  & & ~ \\
  q & r &
\end{pmatrix}$ &
$\begin{pmatrix}
  & s & \\
  s & t & \\
  & & u
\end{pmatrix}$
&&$\to$&
$\begin{pmatrix}
  l & m & \\
  m & n & \\
  & & ~
\end{pmatrix}$
& IH & $\bigcirc$ & $\bigcirc$ \\ \hline
$\begin{pmatrix}
  & p & \\
  & & ~ \\
  q & r &
\end{pmatrix}$ &
$\begin{pmatrix}
  & s & \\
  s & t & \\
  & & u
\end{pmatrix}$
&&$\to$&
$\begin{pmatrix}
  l & m & \\
  m & n &\\
  & & ~
\end{pmatrix}$
& IH & $\bigcirc$ & $\bigcirc$ \\ \hline
$\begin{pmatrix}
  p & q & \\
  & & ~ \\
  r & &
\end{pmatrix}$ &
$\begin{pmatrix}
  & s & \\
  s & t & \\
  & & u
\end{pmatrix}$
&&$\to$&
$\begin{pmatrix}
  l & m & \\
  m & n & \\
  & & ~
\end{pmatrix}$
& IH & $\bigcirc$ & $\bigcirc$ \\ \hline
$\begin{pmatrix}
  p & q & \\
  & & ~ \\
  & r &
\end{pmatrix}$ &
$\begin{pmatrix}
  & s & \\
  s & t & \\
  & & u
\end{pmatrix}$
&&$\to$&
$\begin{pmatrix}
  l & m & \\
  m & n & \\
  & & ~
\end{pmatrix}$
& IH & $\bigcirc$ & $\bigcirc$ \\ \hline\hline
\end{tabular}
\caption{(continued.)}
\end{table}

\newpage

\begin{table}[htbp]
\renewcommand{\arraystretch}{1.05}
\centering
\begin{tabular}{cccccccc} \hline\hline
$M_\nu$ & $M_R$ &&& $M_L$ & spectrum & CP & $0\nu\beta\beta$ \\ \hline
$\begin{pmatrix}
  p & & \\
  & q & \\
  & & ~
\end{pmatrix}$ &
$\begin{pmatrix}
  r & s & \\
  s & t & \\
  & & u
\end{pmatrix}$
&&$\to$&
$\begin{pmatrix}
  l & m & \\
  m & n & \\
  & & ~
\end{pmatrix}$ 
& IH & $\bigcirc$ & $\bigcirc$ \\ \hline
$\begin{pmatrix}
  p & & \\
  & q & \\
  & & ~
\end{pmatrix}$ &
$\begin{pmatrix}
  & & r \\
  & s & t \\
  r & t & u
\end{pmatrix}$
&&$\to$&
$\begin{pmatrix}
  l & m & \\
  m & n & \\
  & & ~
\end{pmatrix}$
& IH & $\bigcirc$ & $\bigcirc$ \\ \hline
$\begin{pmatrix}
  & p & \\
  q & & \\
  & & ~
\end{pmatrix}$ &
$\begin{pmatrix}
  & & r \\
  & s & t \\
  r & t & u
\end{pmatrix}$
&&$\to$&
$\begin{pmatrix}
  l & m & \\
  m & n & \\
  & & ~
\end{pmatrix}$
& IH & $\bigcirc$ & $\bigcirc$ \\ \hline
$\begin{pmatrix}
  p & & \\
  & q & \\
  & & ~
\end{pmatrix}$ &
$\begin{pmatrix}
  & r & s \\
  r & t & \\
  s & & u
\end{pmatrix}$
&&$\to$&
$\begin{pmatrix}
  l & m & \\
  m & n & \\
  & & ~
\end{pmatrix}$
& IH & $\bigcirc$ & $\bigcirc$ \\ \hline
$\begin{pmatrix}
  & p & \\
  q & & \\
  & & ~
\end{pmatrix}$ &
$\begin{pmatrix}
  & r & s \\
  r & t & \\
  s & & u
\end{pmatrix}$ 
&&$\to$&
$\begin{pmatrix}
  l & m & \\
  m & n & \\
  & & ~
\end{pmatrix}$
& IH & $\bigcirc$ & $\bigcirc$ \\ \hline
$\begin{pmatrix}
  p & & \\
  & q & \\
  & & ~
\end{pmatrix}$ &
$\begin{pmatrix}
  & r & s \\
  r & & t \\
  s & t & u
\end{pmatrix}$
&&$\to$&
$\begin{pmatrix}
  l & m & \\
  m & n & \\
  & & ~
\end{pmatrix}$
& IH & $\bigcirc$ & $\bigcirc$ \\ \hline\hline
\end{tabular}
\caption{$M_\nu$ with 7 zeros + $M_R$ with symmetric 2 zeros.}
\end{table}

\clearpage

\end{document}